\documentclass[fleqn,usenatbib]{mnras}

\usepackage{newtxtext,newtxmath}
\usepackage[T1]{fontenc}
\usepackage{ae,aecompl}

\usepackage{graphicx}	% Including figure files
\usepackage{amsmath}	% Advanced maths commands
\usepackage{amssymb}	% Extra maths symbols
\usepackage{subfig}
\usepackage{bm}
\title[Inclination dependence of non-linearity in QPOs]{A Likely Inclination Dependence in the Non-linear Variability of Quasi Periodic Oscillations from Black Hole Binaries}

\author[Arur \& Maccarone.]{
K. Arur,$^{1}$\thanks{E-mail: kavitha.arur@ttu.edu}
T. J. Maccarone,$^{1}$
\\
$^{1}$Department of Physics and Astronomy, Texas Tech University,Lubbock,TX,79409-1051,USA
}

\date{Accepted XXX. Received YYY; in original form ZZZ}

\pubyear{2019}

\begin{document}
\label{firstpage}
\pagerange{\pageref{firstpage}--\pageref{lastpage}}
\maketitle

\begin{abstract}
We present a systematic analysis of the effects of orbital inclination angle on the non-linear variability properties of type-B and type-C QPOs from black hole binaries. We use the bicoherence, a measure of phase coupling at different Fourier frequencies for our analysis. We find that there is a likely inclination dependent change in the non-linear properties of type-C QPOs as the source transitions from a hard intermediate state to a soft intermediate state. High inclination (edge-on) sources show a change from a `web' to a `cross' pattern, while the low inclination (face-on) sources show a change from a `web' to a `hypotenuse' pattern. We present a scenario of a moderate increase in the optical depth of the Comptonising region as a possible explanation of these effects. The bicoherence of type-B QPOs do not exhibit any measurable inclination dependence. 
\end{abstract}

\begin{keywords}
stars: black holes -- methods: statistical -- X-rays: binaries
\end{keywords}

%%%%%%%%%%%%%%%%%%%%%%%%%%%%%%%%%%%%%%%%%%%%%%%%%%

%%%%%%%%%%%%%%%%% BODY OF PAPER %%%%%%%%%%%%%%%%%%

\section{Introduction}

Quasi-Periodic Oscillations (QPOs) are moderately peaked features seen in the power spectra of Black Hole X-ray Binaries (BHXBs). These QPOs are often divided into two broad categories: high frequency ($f$ > 100Hz) and low frequency ($f$<100Hz). Due to the short timescales of the variability, QPOs are powerful probes of the geometry of the inner regions of the accretion disk around the black hole. In this paper, we focus on the Low Frequency QPOs (LFQPOs).

LFQPOs were phenomenologically classified into types A, B and C, as originally identified in XTE~J1550-564 (\citet{Wijnands1999a}, \citet{Remillard2002a}, \citet{Casella2005}). These QPOs have since been observed in several BHXBs. Of these three types, type-C QPOs are the most commonly observed, and are seen in the frequency range of 0.1-30Hz. Type-C QPOs are typically narrow (Q\footnote{Q = $f_{\text{centroid}}$/FWHM} $\geq$ 6), and are observed in the hard intermediate state with large rms variability, often along with higher harmonics. Type-B QPOs, seen in the soft intermediate state, are generally broader (Q $\sim$ 6) and are seen in the 4-7 Hz range. Type-A QPOs are the least commonly observed type. These are seen in the soft state, with their weak and broad (Q $\sim$ 1-3) peaks making them hard to detect.

Various models have been proposed to explain the origin of these LFQPOs. The suggested QPO production mechanisms include models broadly based on two mechanisms: instabilities and geometrical effects. Instability models invoke different processes such as the presence of a transition layer \citep{Titarchuk2004}, propagation of magneto-acoustic waves in the corona \citep{Cabanac2010} or accretion ejection instabiltity \citep{Tagger1999}, such as in \citet{Varniere2002}, where the QPO is produced as a result of spiral density waves produced in the accretion disk due to magnetic stresses. The geometric models usually consider Lense-Thirring precession as proposed by \citep{Stella1998}, where the QPO originates from the relativistic precession of the accretion flow as a solid body \citep{Ingram2009}.

Because these models can all reproduce the observed range of frequencies and amplitudes, sophisticated timing techniques that go beyond the simple power spectrum are required to distinguish between models. One such technique is to characterize and understand the nature of non-linear variability. Non-linearity of the broadband noise in power spectra of accreting black holes has been shown by the presence of a linear rms-flux relation \citep{Uttley2001} and a log-normal flux distribution \citep{Uttley2005}. The bispectrum, a measure of phase coupling among triplets of frequencies, is a higher order time series analysis technique that can be used to break degeneracies between models that reproduce very similar power spectra \citep{Maccarone2005}. The bispectrum has been used to detect the presence of non-linearity in the hard state of Cyg X-1 and GX~339-4 \citep{Maccarone2002}, detect the presence of coupling between broadband noise and QPO frequencies in GRS~1915+105 \citep{Maccarone2011} and to study the evolution of the type-C QPOs across state transitions in GX~339-4 \citep{Arur2019}. 

Inclination dependence of various properties of QPOs have been reported previously. \citet{Motta2015} found that the QPO amplitude is dependent on the orbital inclination angle, where the type C QPOs are stronger in high inclination (edge-on) systems, and type B QPOs are stronger in low inclination (face-on) systems. 

\citet{Heil2015} showed that the effect of QPOs on the power-colour\footnote{ratios of integrated power in different Fourier frequency ranges} properties is inclination dependent. Additionally, it was found by \citet{VandenEijnden2017} that both the evolution and sign of the energy dependent phase lags of type C QPOs show a strong dependence on inclination. Hints of an inclination dependence was also seen by \citet{DeRuiter2019} on the evolution of the phase difference between the QPO and its harmonic, with high inclination sources showing a more consistent pattern than the low inclination ones. 

In this paper, we report on the inclination dependence of the non-linear variability seen from BHXBs using archival data from the \textit{X-ray Timing Explorer (RXTE)}/Proportional Counter Array (PCA). 

In Section~\ref{sec:stat} we give a brief overview of the statistical methods used in this paper, followed by a description of the data sample and analysis in Section~\ref{sec:data}. In Section~\ref{sec:results}, we present the results on the evolution of the bicoherence patterns of QPOs from low and high inclination sources, and then discuss these results in Section~\ref{sec:discussion}. Finally, we present our conclusions in Section~\ref{sec:conclusions}. 

\section{Statistical Methods}
\label{sec:stat}

\subsection{Bicoherence}

The bispectrum is a higher order time series analysis technique that can be used to study nonlinear interactions via the coupling of phases of Fourier components. In the same way that the power spectrum is the Fourier domain equivalent of the 2-point correlation function, the bispectrum is the Fourier domain equivalent of the 3-point correlation function. The bispectrum of a time series divided into K segments is given by:

\begin{equation}
    B(k,l) =\frac{1}{K} \sum_{i=0}^{K-1}X_i(k)X_i(l)X^*_i(k+l)
	\label{eq:bispectrum}
\end{equation}

where $X_i(f)$ is the Fourier transform of the $i$th segment of the time series at frequency f, and $X^*_i(f)$ is the complex conjugate of $X_i(f)$.  

A related term is the bicoherence, which is given by: 

\begin{equation}
    b^2 =\frac{\left|\sum X_i(k)X_i(l)X^*_i(k+l) \right|^2}
    {\sum|X_i(k)X_i(l)|^2 \sum |X_i(k+l)|^2} 
	\label{eq:bicoherence}
\end{equation}

This normalization, proposed by \citet{Kim1979}, measures the fraction of power at frequency $k+1$ due to the coupling of the three frequencies. The value of the bicoherence will have a value of 1 if the phase of the bispectrum (biphase) remains constant over time, and approaches zero for large number of measurements if the phases are random. Since the bicoherence is normalised to lie between 0 and 1, a bias of $1/K$ is subtracted from the bicoherence measurement. 

The expectation value of the bispectrum is unaffected by Gaussian noise, but the variance is higher in noisy signals. However, due to the nonlinear nature of Poisson noise, it can strongly effect the bicoherence at frequencies where the Poisson noise level is comparable to that of the intrinsic variability \citep{Uttley2005}. In this paper, we study the bicoherence of QPOs which have low frequencies and have high rms variability, and thus are not strongly affected by the effects of Poisson noise.

\subsection{Biphase}
\label{sec:biphase}

The bispectrum, being a complex number, can be represented as a combination of a magnitude and a phase. This phase is called the biphase. The biphase is defined over the full 2$\pi$ interval, and contains valuable information about the shape of the underlying waveform. Since there always exists a biphase, it must be noted that the value of the biphase is only meaningful in the regions of statistically significant bicoherence. 

The biphase of $f_1, f_2$ and $f_{1+2}$ gives the average phase difference between these three components. Thus the biphase where both $f_1$ and $f_2$ are equal to $f_{QPO}$ gives the phase difference between the fundamental and first harmonic of the QPO. Using this method, the phase difference between any two harmonics is easily calculated.

The imaginary component of the bispectrum contains information on the reversibility of the lightcurve, while the real component probes the asymmetry of the underlying flux distribution. Thus certain aspects of non-linearity (such as the kurtosis of the flux distribution) cannot be revealed using the bispectrum, and require the use of further higher order spectra. For a more detailed overview of the biphase, see \citet{Maccarone2013}.

\section{Observations and Data Reduction}
\label{sec:data}
\subsection{Data Sample}

For our analysis, we use the sample of archival \textit{RXTE} observations of 14 Galactic black hole binaries analysed in \cite{Motta2015}. A complete list of the observations that were analysed can be found as supplementary material in the online version of this paper. 

For 9 of these sources, either a direct estimation of the inclination angle or an upper or lower limit was obtained. The details of the inclination angles of the sources are listed in Table~\ref{table:sample}. Apart from these estimates, the presence or absence of absorption dips was used as the main discriminator between high and low inclination sources respectively.

The sources with inclination angles greater than $\approx$ 65-70 degrees are classified as high inclination, and low inclination below this value. XTE J1859+226 and MAXI J1543-564 were classified as intermediate inclination as they were not unambiguously a part of the previous two categories. (For more details on the classification of the inclination angle of these sources, see Section 2 and Appendix B of \citet{Motta2015}).

For a reliable estimation of the bicoherence, it is necessary that the observation has a high count rate, as well as is long enough to observe multiple cycles at the frequency of interest. For this reason, only observations that have a count rate of >750 counts/s/PCU and an observation length of >1ks are included in our sample. Additionally, observations showing significant (>25\%) variations in the mean value of the flux between segments were excluded, as they are considered to be non-stationary.

\begin{table*}
\caption{List of sources and outbursts used for analysis in this work. [1]\citet{Neustroev2014} [2] [3]\citet{Orosz2004} [4] \citet{Munoz-Darias2008a} [5] \citet{Miller-Jones2011} [6] \citet{Corral-Santana2011} [7] \citet{Orosz2011} [8]\citet{Greene2001} [9]\citet{Steiner2011}}
\centering
	\begin{tabular}{|l c c c c|}
    \hline
	Source & Comment & Outburst & Inclination & Reference  \\ 
		&  & & Angle & \\ 
    \hline
    Swift J1735.5-01 & Low & 2005-2010 & $\sim$40-55$^\circ$ & [1]\\
    4U~1543-47 & Low & 2002 & 20.7 $\pm$ 1.5$^\circ$ & [2] \\
    XTE~J1650-500 & Low & 2001 & >47$^\circ$ & [3] \\
 	GX~339-4 & Low & 2002,2004,2007,2010& $\geq$40$^\circ$ & [4] \\
 	XTE~J1752-223 & Low & 2009 & $\leq$ 49$^\circ$ & [5]\\
    XTE~J1817-330 & Low & 2006\\
    
    XTE~J1859+226 & Int & 1999 & $\geq$60$^\circ$ & [6] \\
    MAXI~J1543-564 & Int & 2011 &  \\
    
    XTE~J1550-564 & High & 1998, 2000 & 74.7 $\pm$ 3.8$^\circ$ & [7]\\
    4U~1630-40 & High & 2002-03 & \\
    GRO~J1655-40 & High & 2005 & 70.2 $\pm$ 1$^\circ$ & [8] \\
    H1743-322 & High & 2003,2004,2008,2009,2010 & 75 $\pm$ 3$^\circ$ & [9]\\
    MAXI~J1659-152 & High & 2010 & \\
    XTE~J1748-288 & High & 1998 & \\

    \hline
	\end{tabular}
\label{table:sample}
\end{table*}

\subsection{Data Reduction and Analysis}

We examine the timing data of the observations using Single Bit, Event or GoodXenon mode data. The data were filtered to exclude periods of high offset, Earth occultation and passage through the South Atlantic Anomaly (SAA). Data from the 2-30keV energy band was extracted for analysis. 

We obtained background corrected \textsc{Standard 2} data in the bands B = 3.6-6.1 keV (corresponding to Std2 channels 4-10 in epoch 5) and C= 6.5-10.2 keV (corresponding to Std2 channels 11-20). The hardness ratio is then defined to be HR = C/B \citep{Homan2005a,Motta2015}.

In order to extract the power spectra, the data from each observation were binned with a time resolution of 1/128s (with the exception of XTE J1748-288, where a binning of 1/256s was used due to a higher QPO frequency). The data are Fourier transformed using 8s or 16s segments and averaged. The power spectra were then normalised according to \citet{Leahy1983}.  The Poisson noise is not subtracted in the calculation of the bicoherence, so for the regions of the power spectrum at the highest frequencies, the bicoherence may be suppressed.  In practice, this is not important for the purposes of this paper because of the aforementioned focus on high count rate observations, and on the lower frequency parts of the power spectrum. The bicoherence and biphase of the observations were calculated using the equations shown in Section~\ref{sec:stat}.

\section{Results}
\label{sec:results}

\subsection{Bicoherence Patterns}

For the classification of the different phenomenological patterns seen in the bicoherence, we follow the conventions used in \cite{Maccarone2011} and \cite{Arur2019}. 

\subsubsection{Hypotenuse}
\begin{figure} 
	\includegraphics[width=\columnwidth]{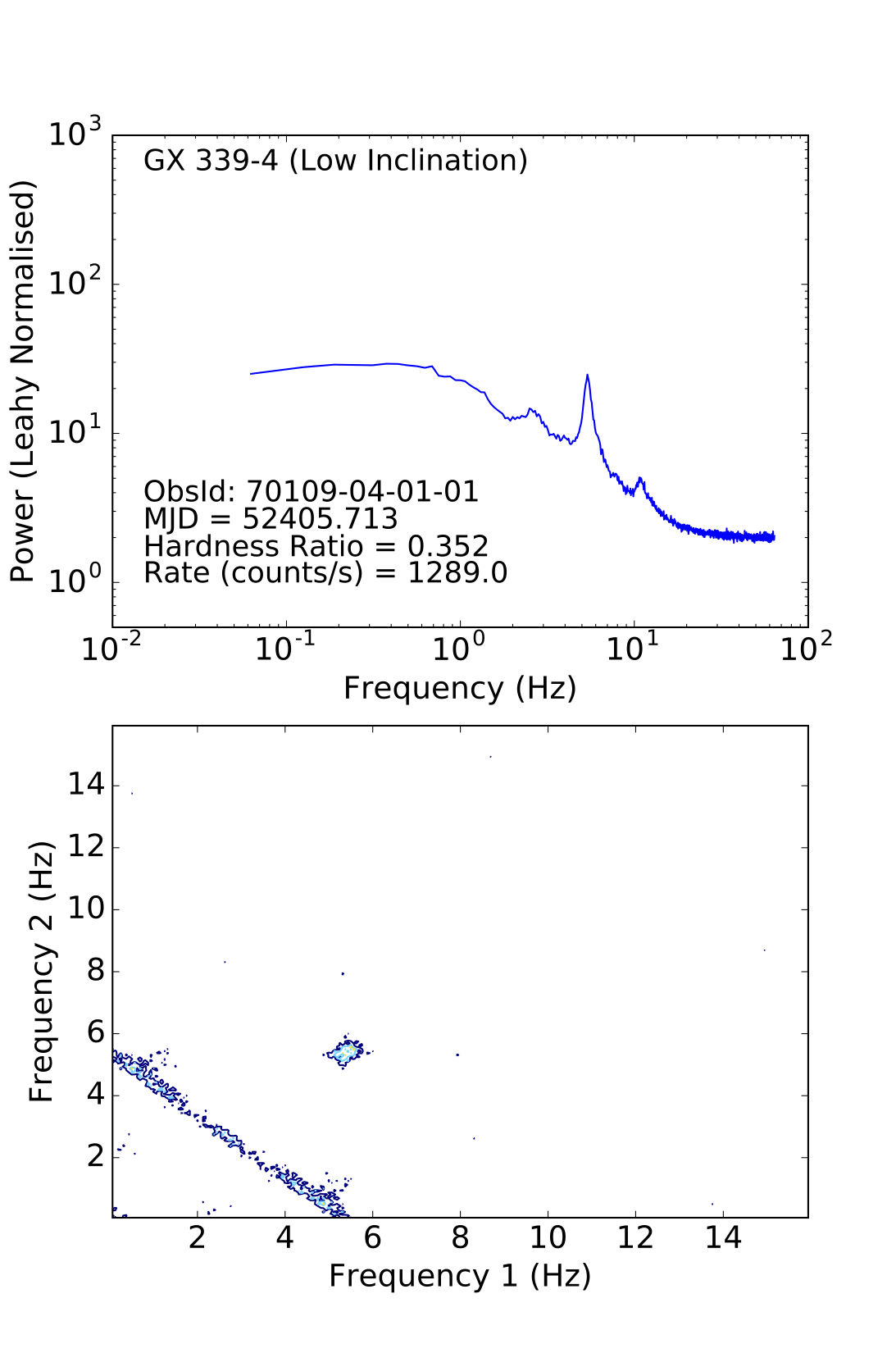}
    \caption{The power spectrum (Top panel) and the bicoherence plot (Bottom panel) showing the 'hypotenuse' pattern from the low inclination source GX 339-4 (observation 70109-04-01-01). The colour scheme of log$b^2$ is as follows: dark blue:-2.0, light blue:-1.75, yellow:-1.50, red:-1.25}
    \label{fig:hyp}
\end{figure}

The `hypotenuse' pattern is seen where a high bicoherence is seen in the diagonal region where the two frequency components add up to the frequency of the QPO (see Fig.~\ref{fig:hyp}). This pattern also shows a region of high bicoherence where both $f_1$ and $f_2$ are equal to $f_{QPO}$. This indicates coupling between the fundamental and harmonic frequency components. This pattern is seen in predominantly the low inclination sources in our sample.

\subsubsection{Cross}

\begin{figure}
	\includegraphics[width=\columnwidth]{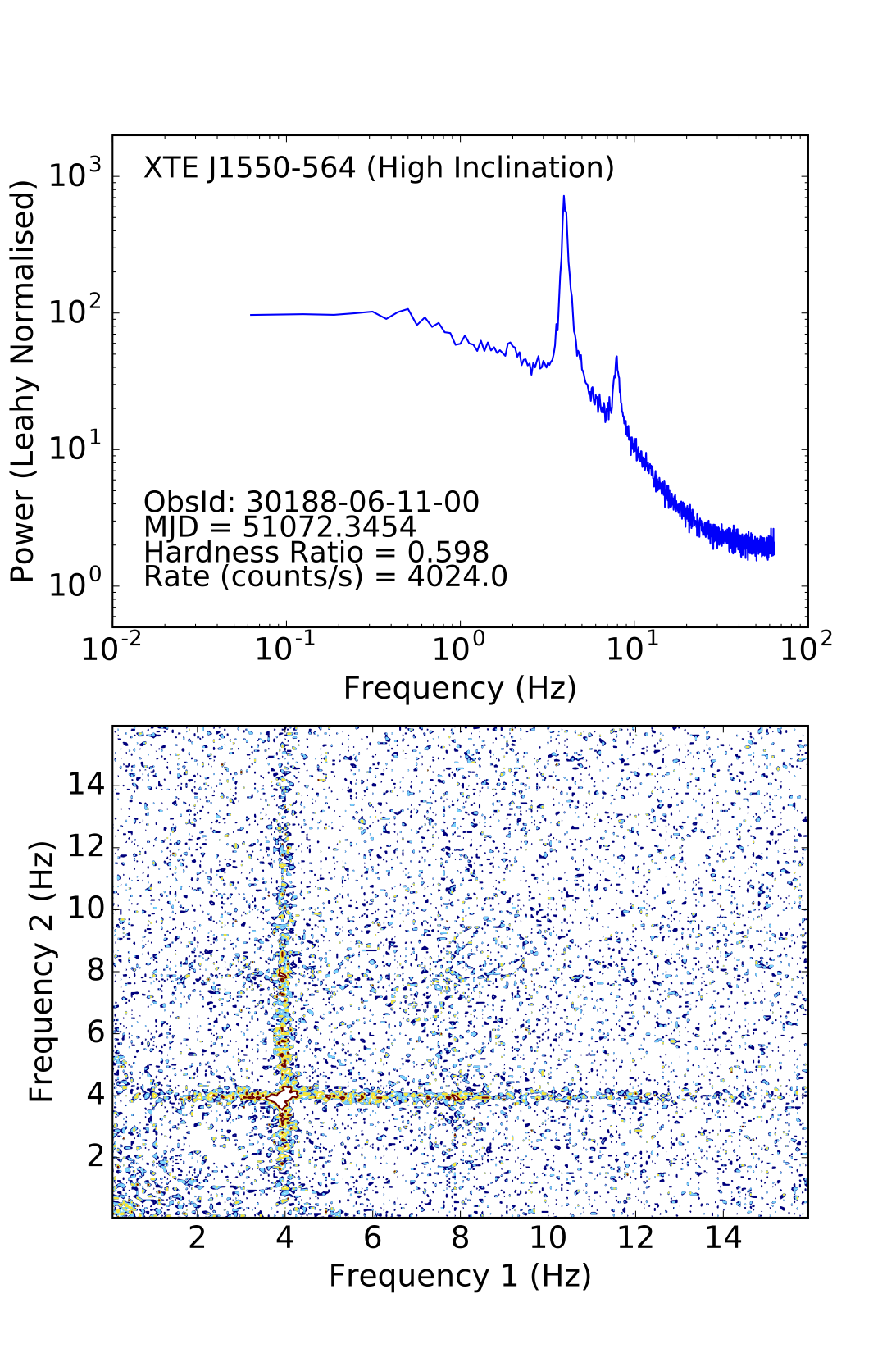}
    \caption{The power spectrum (Top panel) and the bicoherence plot (Bottom panel) showing the 'cross' pattern from the high inclination source XTE J1550-564 (observation 30188-06-11-00). The colour scheme of log$b^2$ is as follows: dark blue:-2.0, light blue:-1.75, yellow:-1.50, red:-1.25}
    \label{fig:cross}
\end{figure}

In the `cross' pattern, high bicoherence is seen for frequency pairs where one frequency is that of the QPO, and the other frequency can be of any value. This leads to the prominent vertical and horizontal streaks as seen in Fig.~\ref{fig:cross}. This pattern is seen in predominantly the high inclination sources in our sample. 

\subsubsection{Web}

\begin{figure}
	\includegraphics[width=\columnwidth]{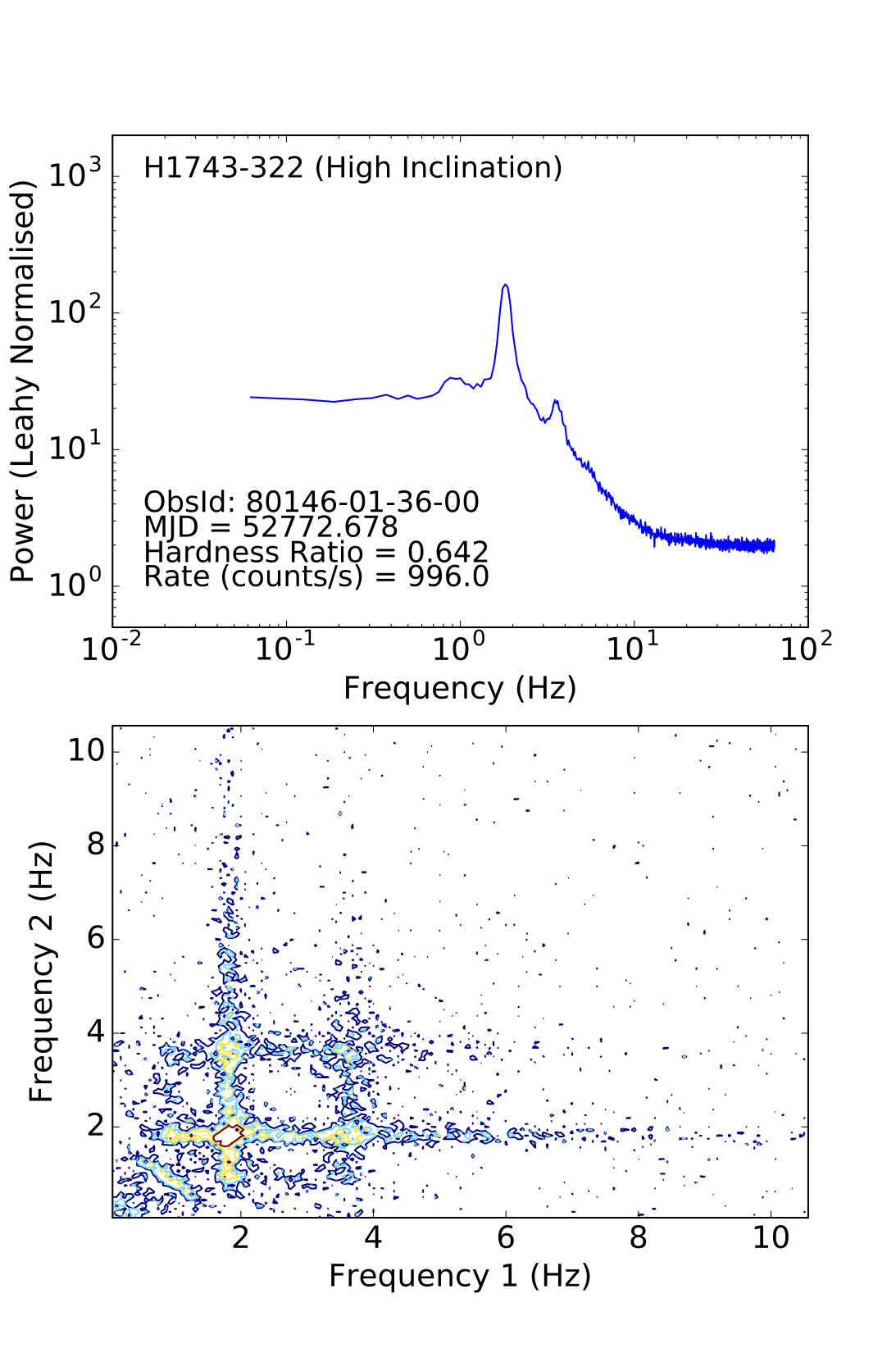}
    \caption{The power spectrum (Top panel) and bicoherence plot (Bottom panel) showing the 'web' pattern from the high inclination source H1743-322 (observation 80146-01-36-00). The colour scheme of log$b^2$ is as follows: dark blue:-2.0, light blue:-1.75, yellow:-1.50, red:-1.25}
    \label{fig:web}
\end{figure}

The `web' pattern is a hybrid class, where both the diagonal feature of the `hypotenuse' and the vertical and horizontal streaks from the `cross' pattern are seen. An example of this pattern can be seen in Fig.~\ref{fig:web}.  This pattern is seen in both the high and low inclination sources in our sample, mostly in observations where the QPO frequency is low (< 2 Hz) and the source is in LHS/HIMS. 

\subsection{Evolution of type C QPOs}
\subsubsection{Low inclination sources}

\begin{figure*}
  \centering
  \subfloat{\includegraphics[width=0.75\columnwidth]{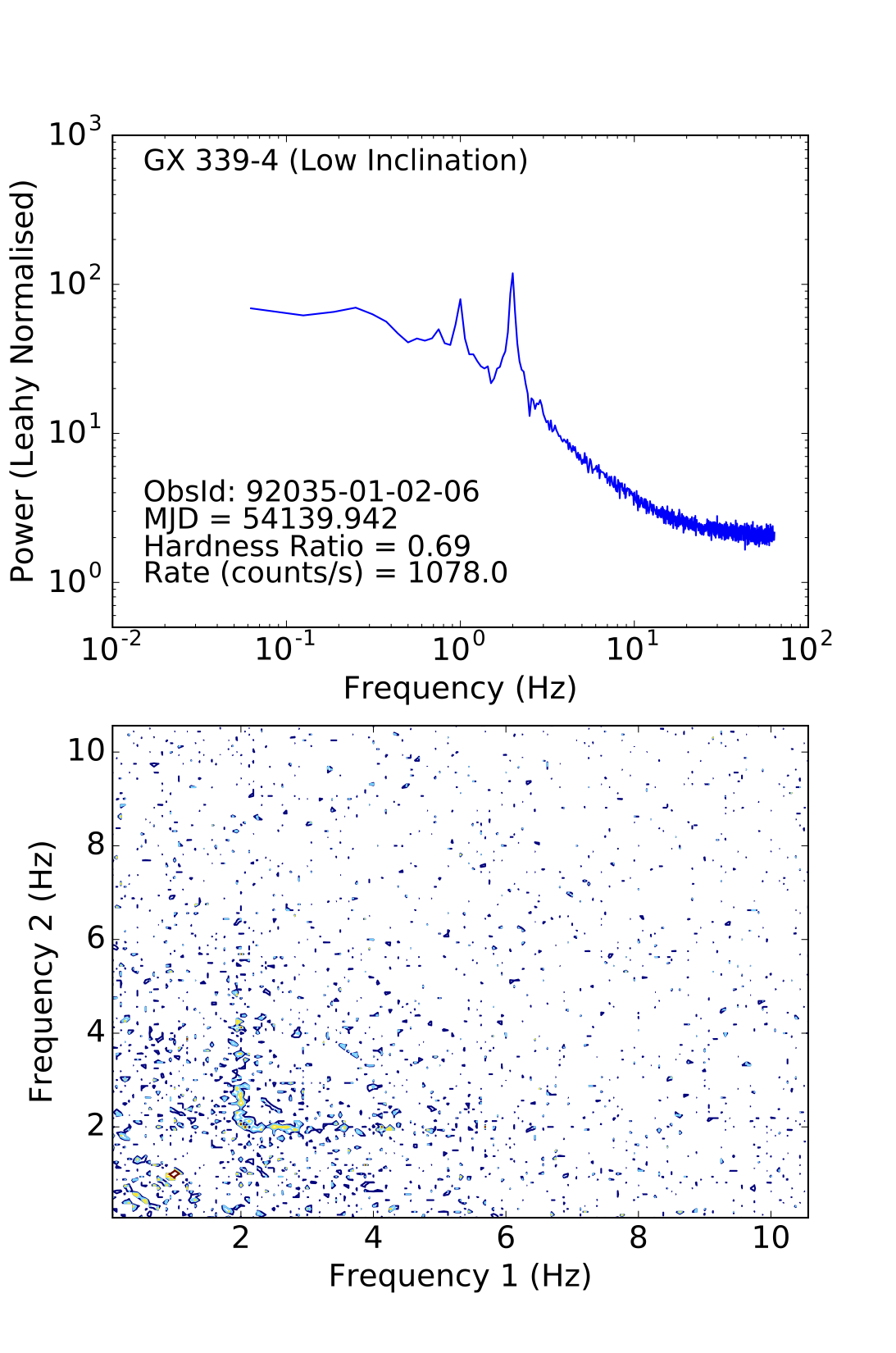}}
  \subfloat{\includegraphics[width=0.75\columnwidth]{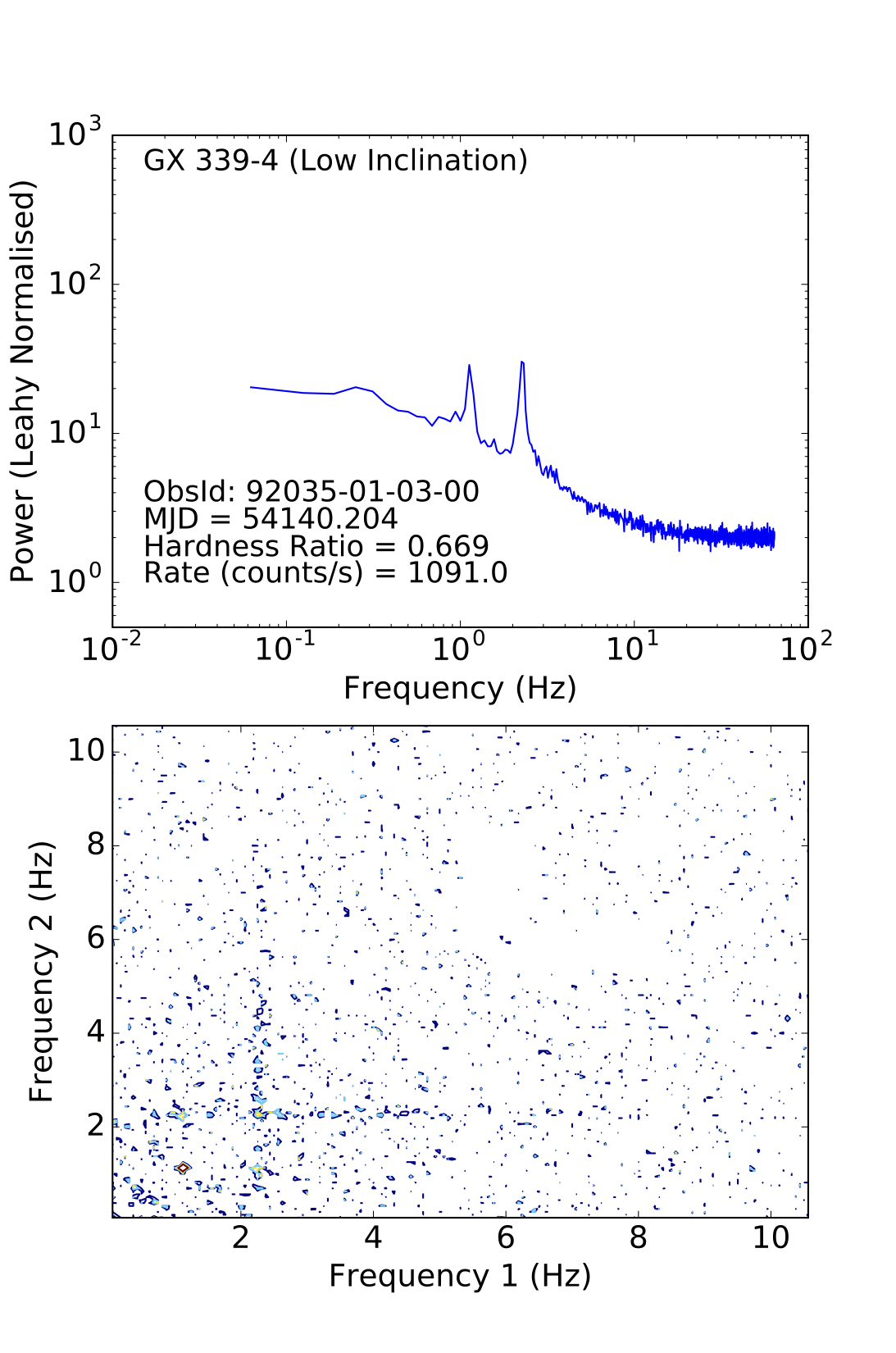}}
  \subfloat{\includegraphics[width=0.75\columnwidth]{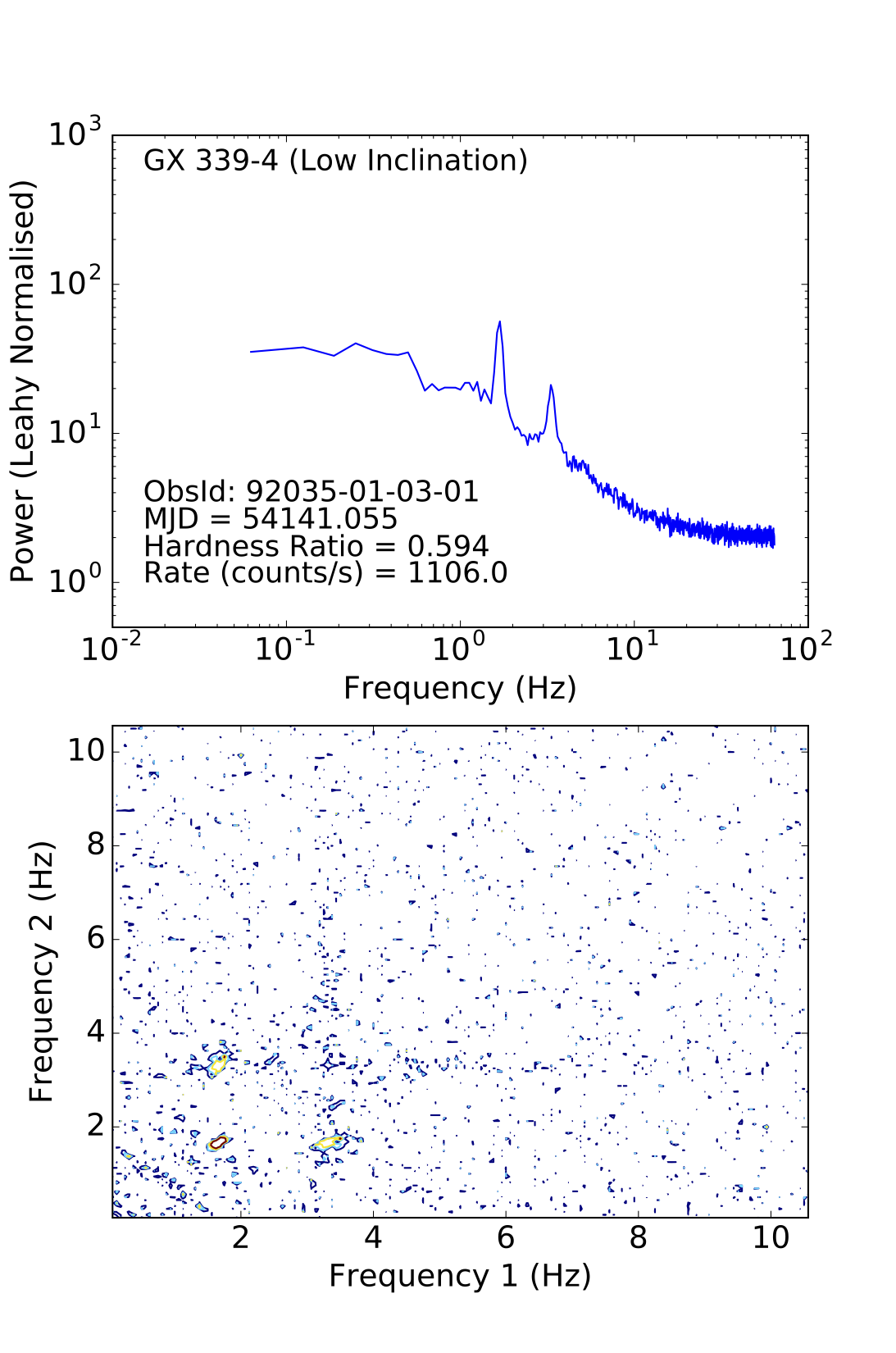}}\\
  \subfloat{\includegraphics[width=0.75\columnwidth]{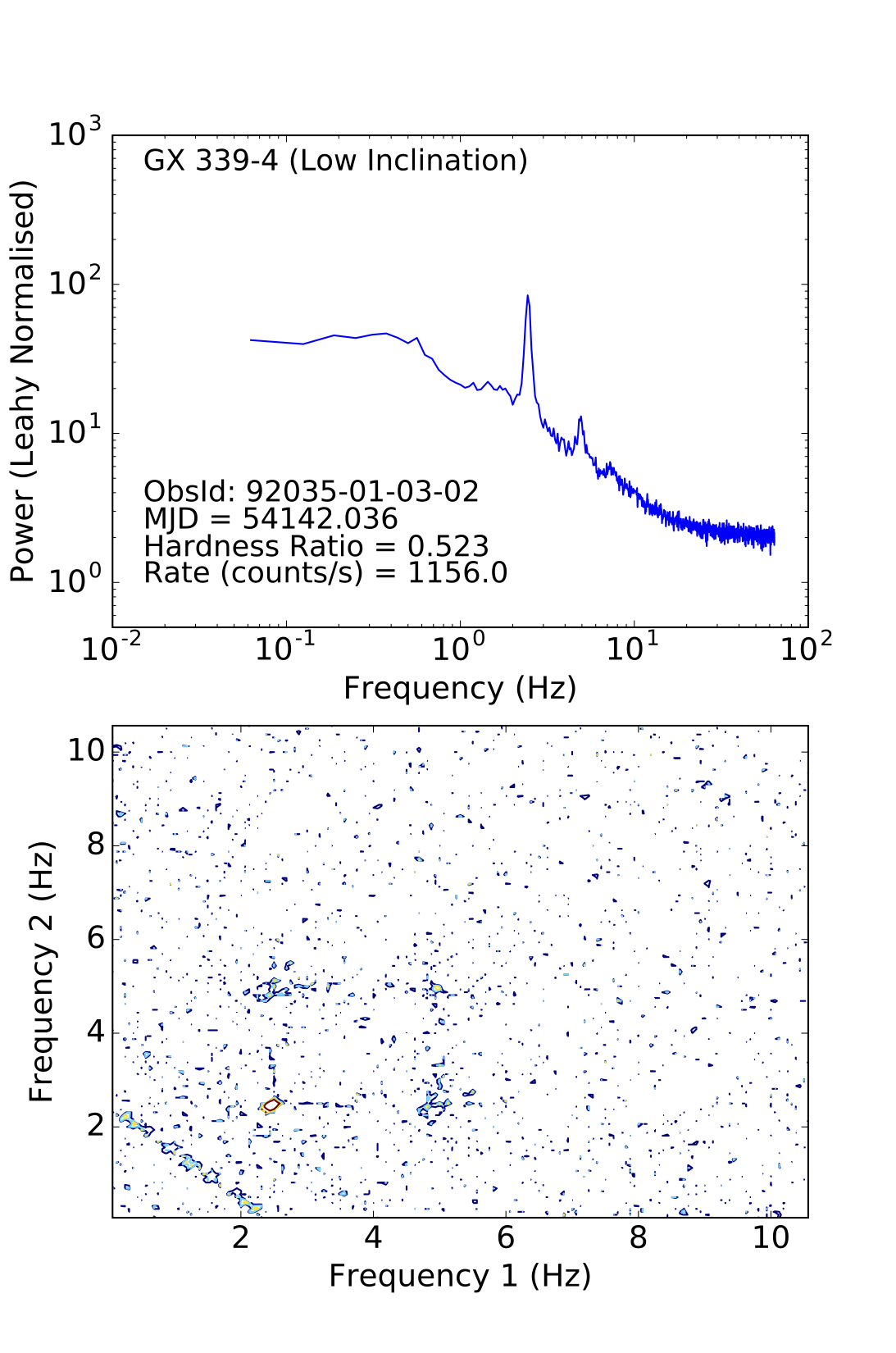}}
  \subfloat{\includegraphics[width=0.75\columnwidth]{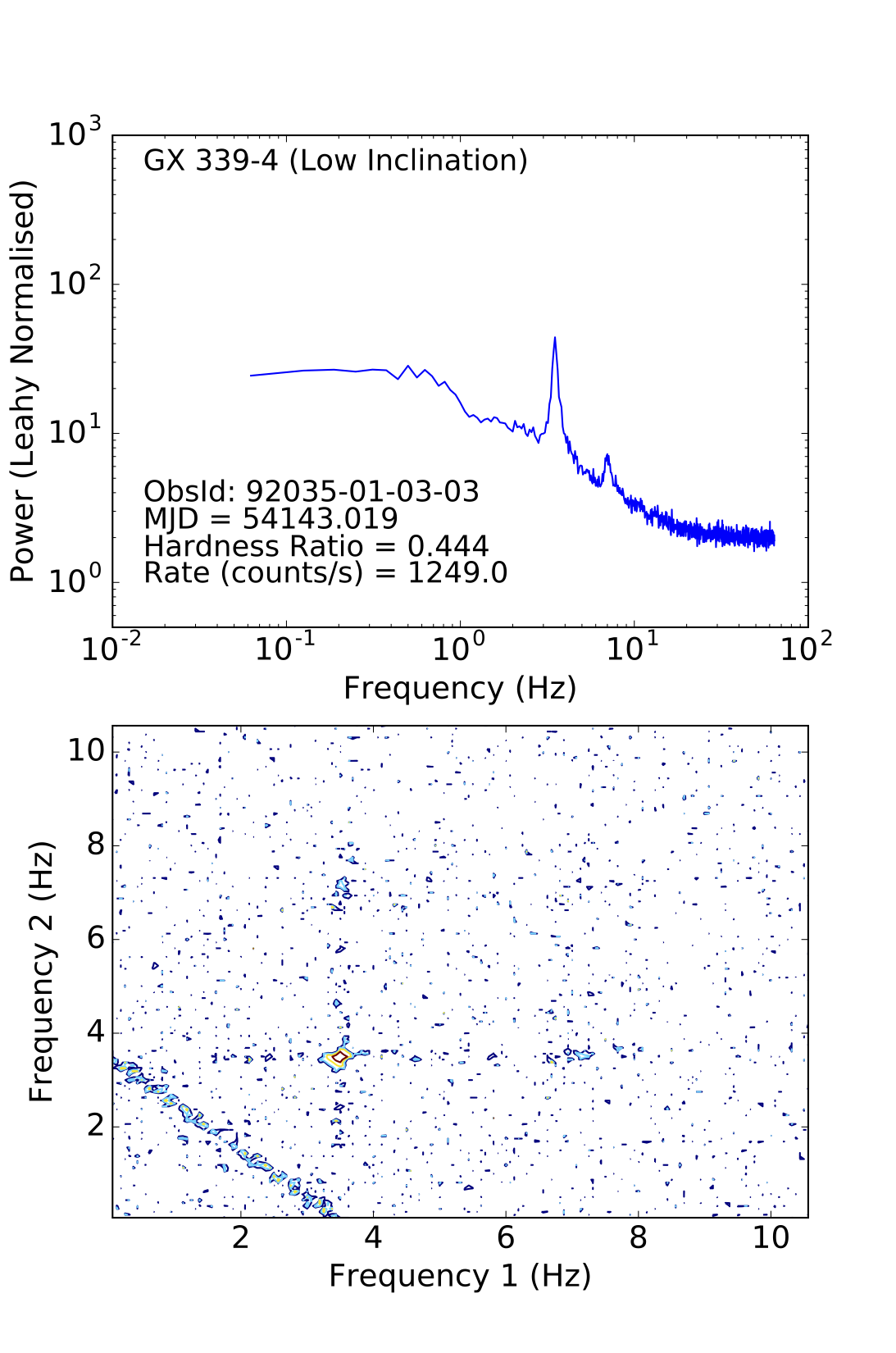}}
  \subfloat{\includegraphics[width=0.75\columnwidth]{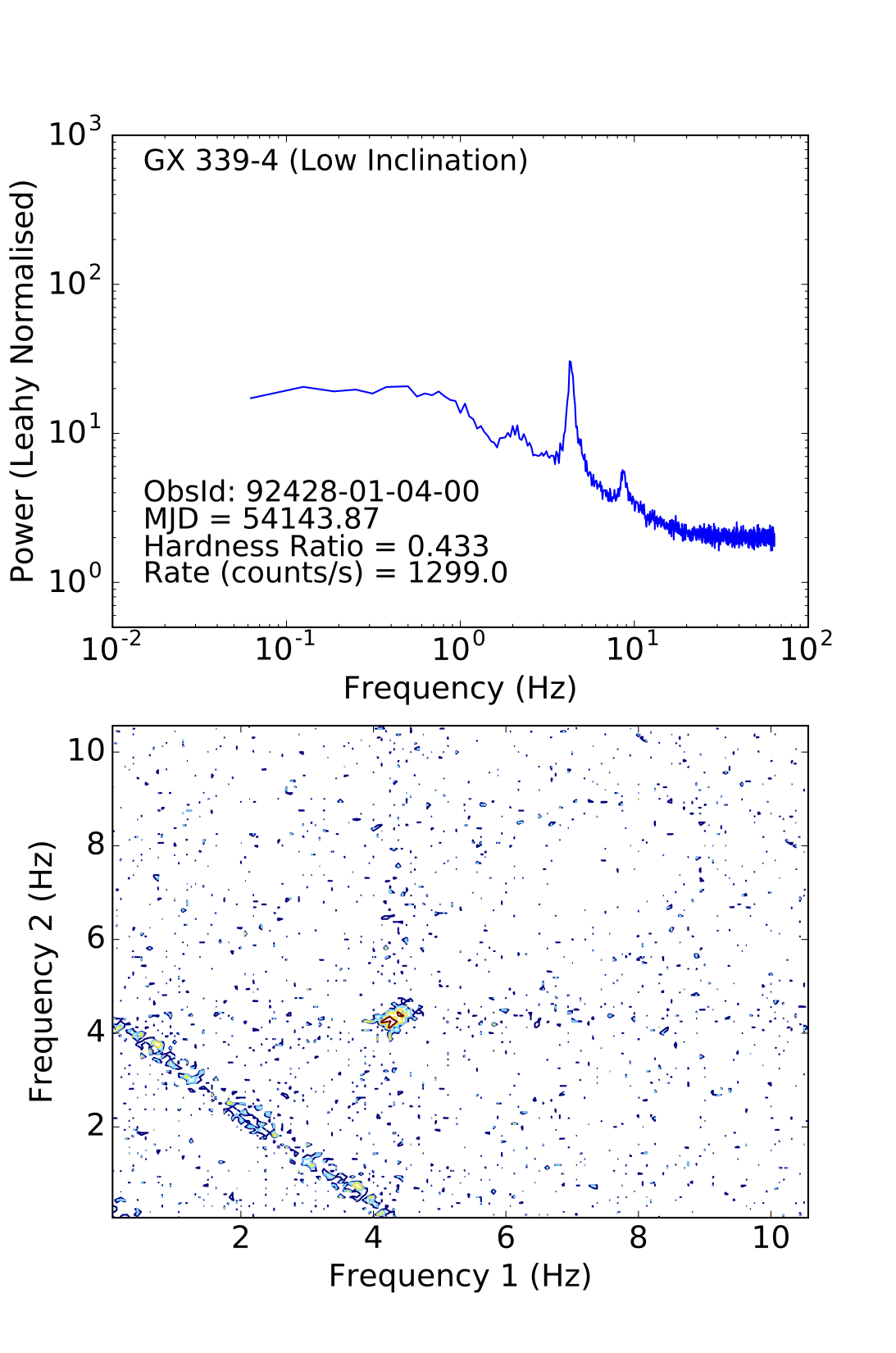}}
  \caption{The power spectrum and the bicoherence plot for multiple observations of GX~339-4 during the 2007 outburst. The colour scheme of log$b^2$ is as follows: dark blue:-2.0, light blue:-1.75, yellow:-1.50, red:-1.25. As the source evolves from HIMS to SIMS, the bicoherence pattern gradually evolves from a `web' to a `hypotenuse' pattern. The colour version of this figure can be found in the online version of this paper.}
  \label{fig:li_evolution}
\end{figure*}

It has previously been shown in \cite{Arur2019} that during the softening phase of the outburst, as GX 339-4 moves from HIMS to SIMS, a gradual change is seen in the bicoherence. The pattern moves from being a `web' to a `hypotenuse', as the strength of the bicoherence along the diagonal steadily increases, and that along the vertical and horizontal streaks falls off. This is shown in Fig~\ref{fig:li_evolution} for the 2007 outburst where this change is seen most clearly. This transition is also observed during the outbursts that occurred in 2002 and 2010. Due to fact that only a few observations are available during the 2004 outburst, we do not observe any change in the bicoherence in this period.

Three of the six low inclination sources in our sample\footnote{As we expand our study of the non-linear properties of QPOs to multiple sources, it must be noted that our sample of type C QPOs from low inclination sources is dominated by GX~339-4.} only have a few observations of QPOs during state transition. From XTE~J1752-223, two out of the three observed type C QPOs show the `hypotenuse' before a type B QPO is observed. There are only 2 observations of type C QPOs from XTE~J1817-330, both with QPO rms < 3\%.  Similarly, the three observations of Type C QPOs from 4U~1543-47 before the source transitions into SIMS also have very low (< 3\%) QPO rms. 

Swift~J1735.5-01 entered the HIMS, but instead of transitioning into the soft states, returned to the hard state after a few months. Thus the data from this source are dominated by type C QPOs with frequencies $f$<1Hz. 

XTE~J1650-500 is the only other low inclination source in our sample that has significant coverage of during the 2001 outburst. During this outburst, a transition from `web' to `hypotenuse' is observed, but is not as clearly visible as GX~339-4.

\subsubsection{High inclination sources}

\begin{figure*}
  \centering
  \subfloat{\includegraphics[width=0.75\columnwidth]{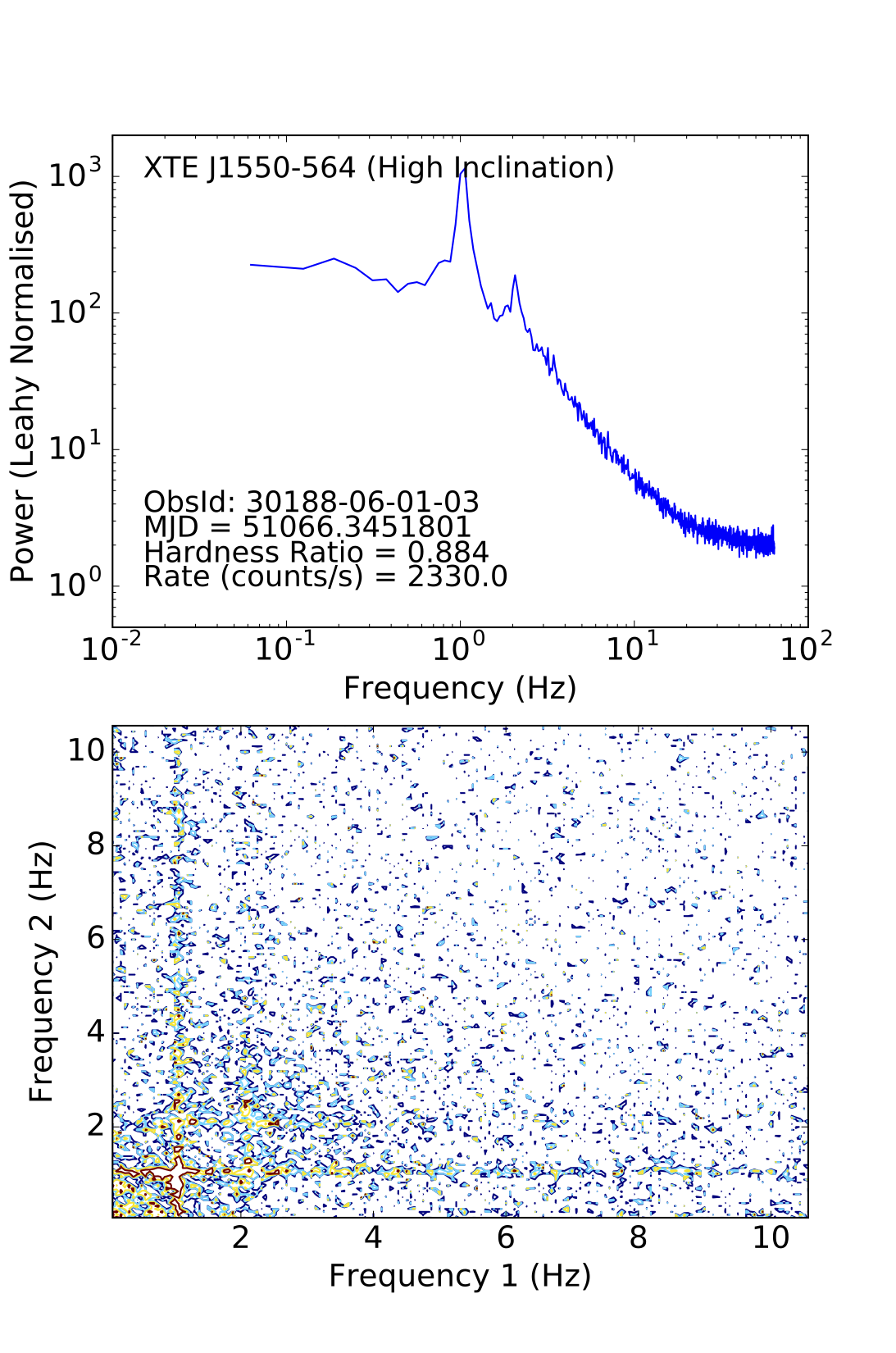}}
  \subfloat{\includegraphics[width=0.75\columnwidth]{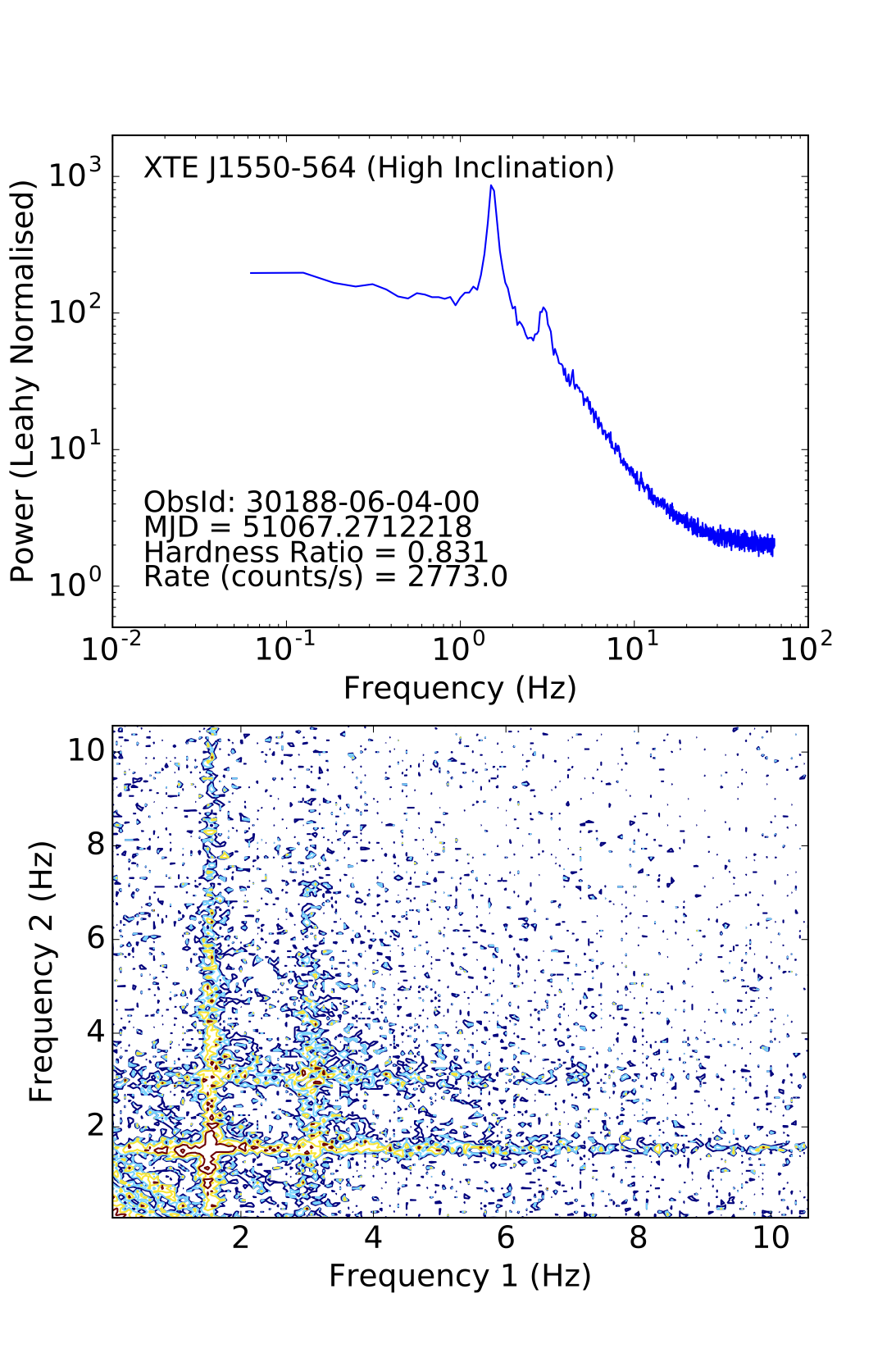}}
  \subfloat{\includegraphics[width=0.75\columnwidth]{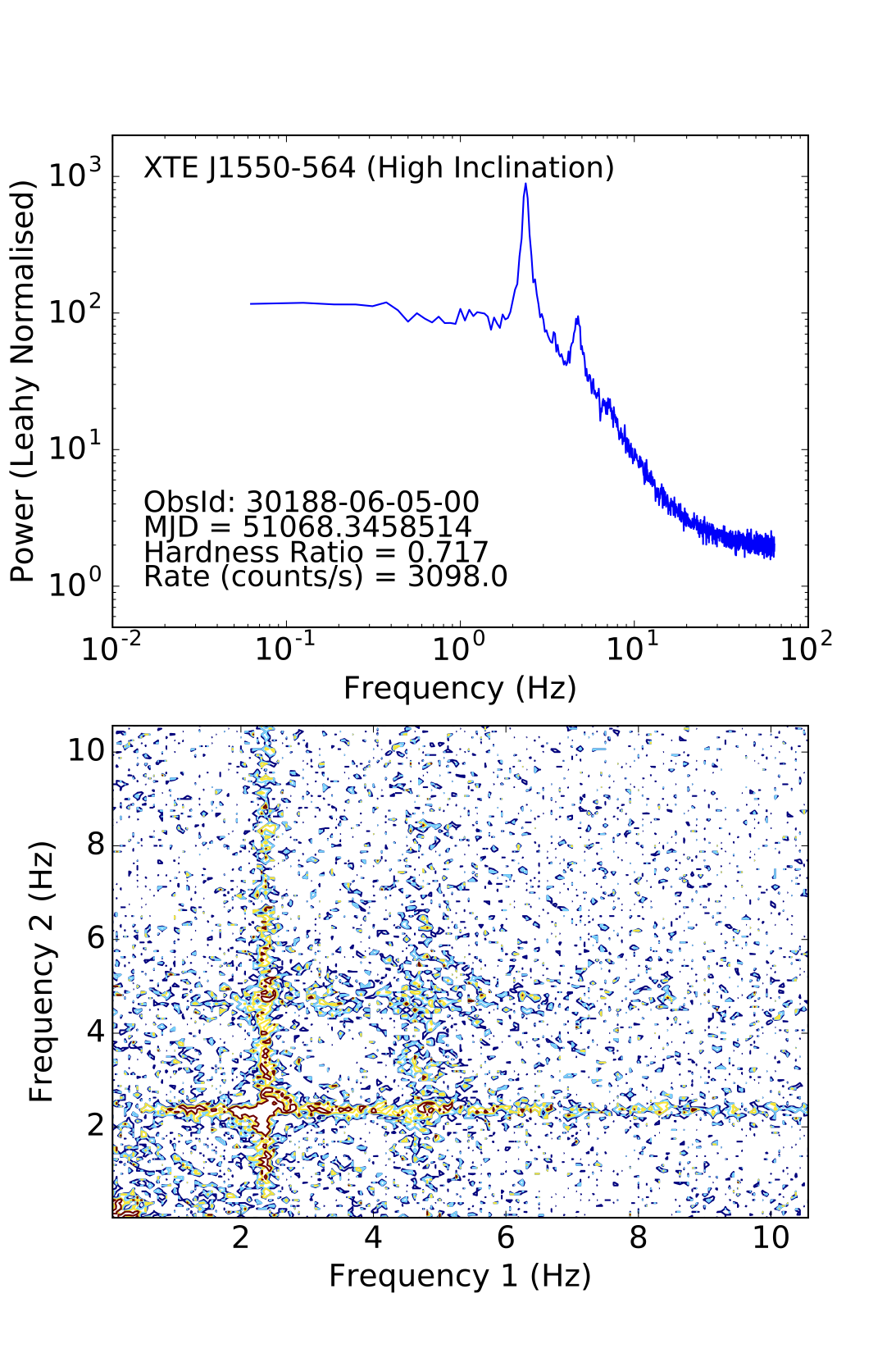}}\\
  \subfloat{\includegraphics[width=0.75\columnwidth]{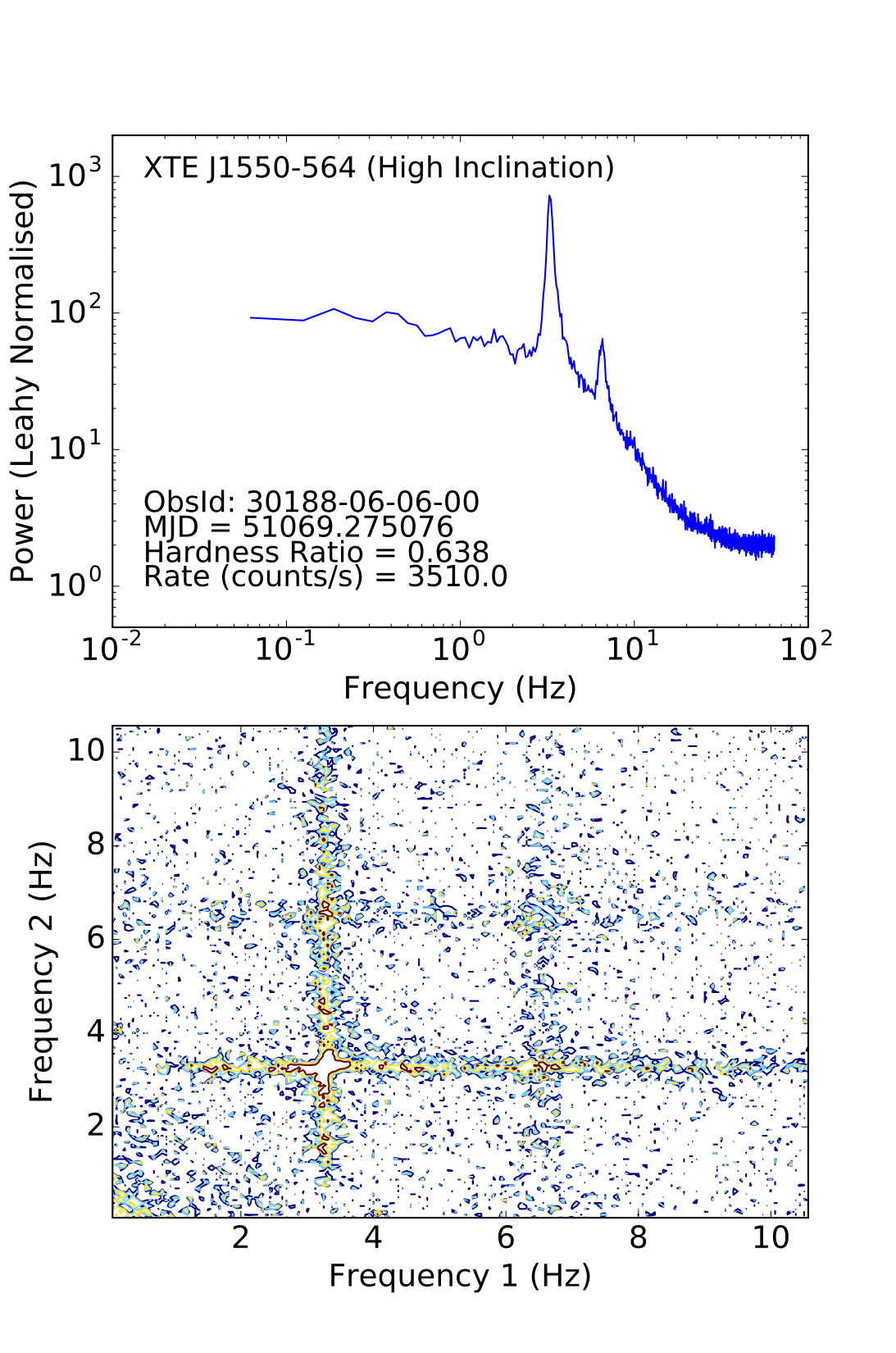}}
  \subfloat{\includegraphics[width=0.75\columnwidth]{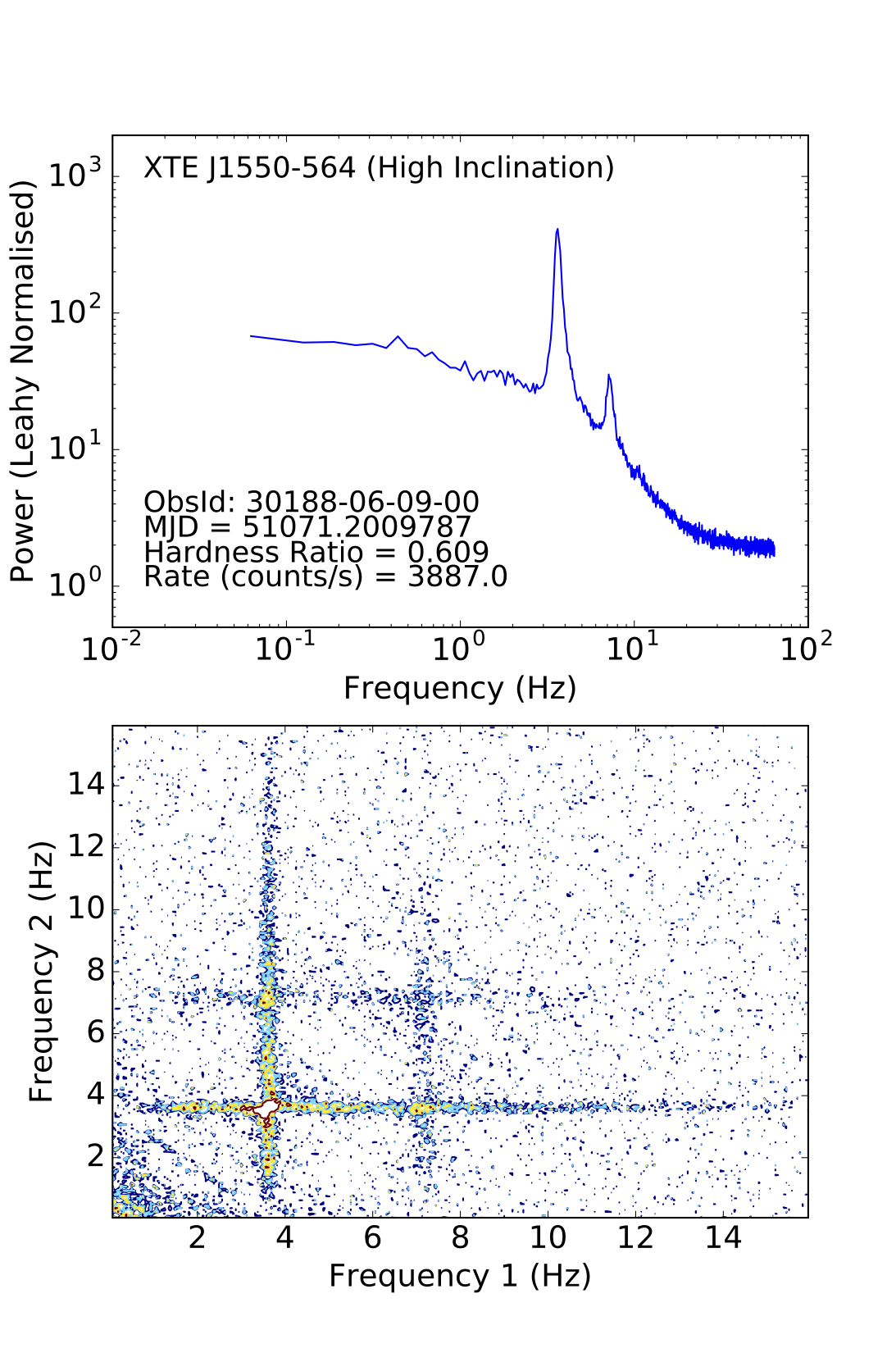}}
  \subfloat{\includegraphics[width=0.75\columnwidth]{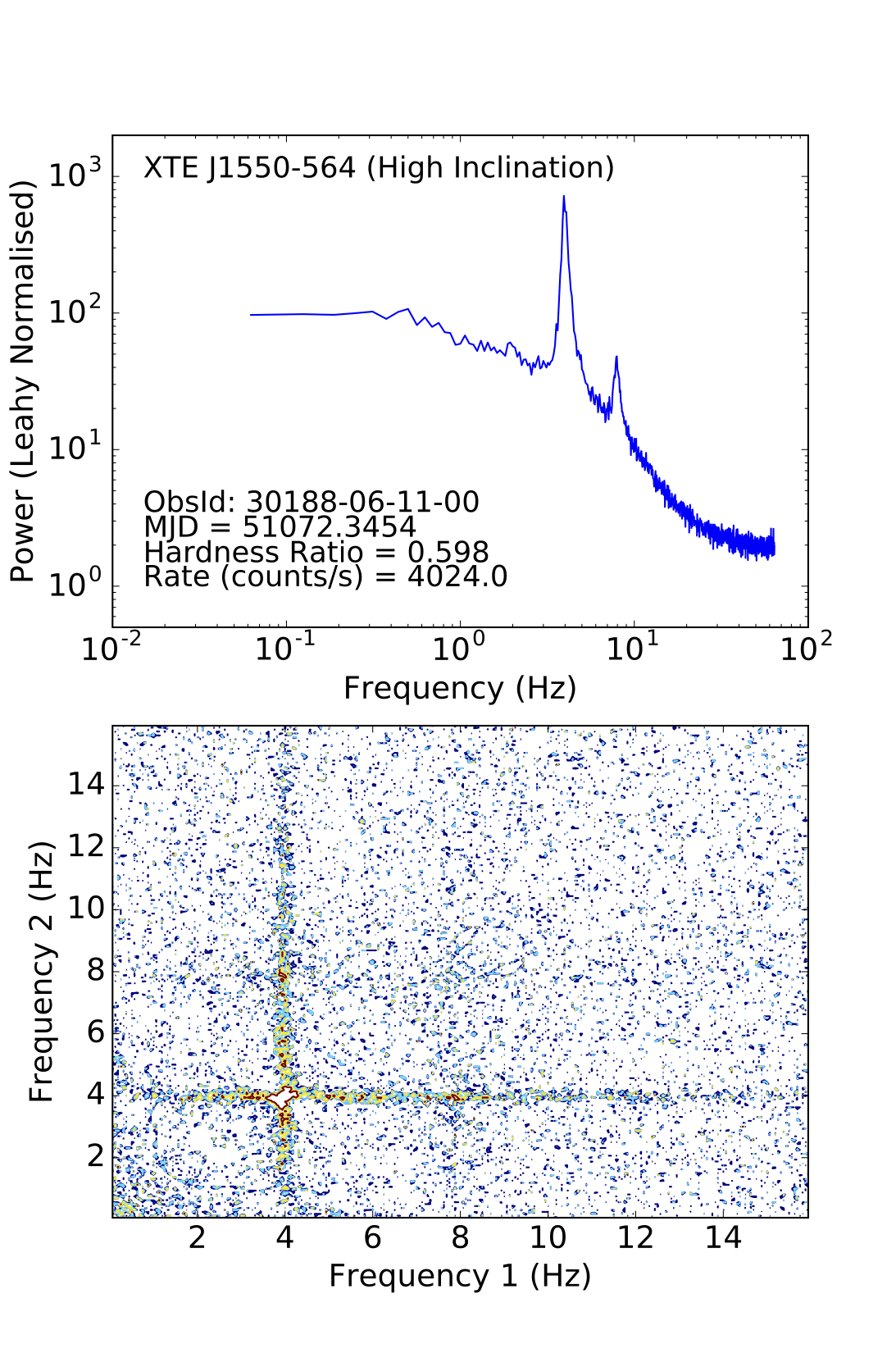}}
  \caption{The power spectrum and the bicoherence plot for multiple observations of XTE~J1550-564 during the outburst in 1998. The colour scheme of log$b^2$ is as follows: dark blue:-2.0, light blue:-1.75, yellow:-1.50, red:-1.25. As the source evolves from HIMS to SIMS, the bicoherence pattern gradually evolves from a `web' to a `cross' pattern. The colour version of this figure can be found in the online version of this paper. }
  \label{fig:hi_evolution}
\end{figure*}

In the sample of high inclination sources, we observe a consistent  pattern of change through the state transition in the bicoherence from XTE~J1550-564, GRO~J1655-40, H~1743-322 and MAXI~J1659-152. In these objects, as the source transitions from the HIMS towards the SIMS, the strength of the bicoherence along the `hypotenuse' decreases, while the strength of the bicoherence along the vertical and horizontal streaks increases. This leads to a gradual transition from a `web' pattern to a `cross' pattern. Fig~\ref{fig:hi_evolution} shows an example of such a transition as seen in XTE~J1550-564.  

No patterns were detected in the bicoherence of 4U~1630-47, but this may be due to poor sensitivity. For this source, almost all the observations had a QPO rms<3\%. The 2 observations that have a high QPO rms both have count rates < 100 counts/sec. 

\subsubsection{Hardening phase of the outburst}

When the source is in the HIMS during the hardening phase of the outburst, 3 of the high inclination sources [GRO~J1655-40, H1743-322 and XTE~J1748-288] briefly show the `hypotenuse' pattern in their bicoherence. However, due to either a lack of further observations (in the case of GRO~J1655-40) or a re-softening of the source (H1743-322), the evolution of this pattern could not be analysed. In the case of XTE~J1748-288, only one observation showed the `hypotenuse' pattern, with all subsequent observations exhibiting no detectable pattern in their bicoherence.

\begin{figure}
	\includegraphics[width=\columnwidth]{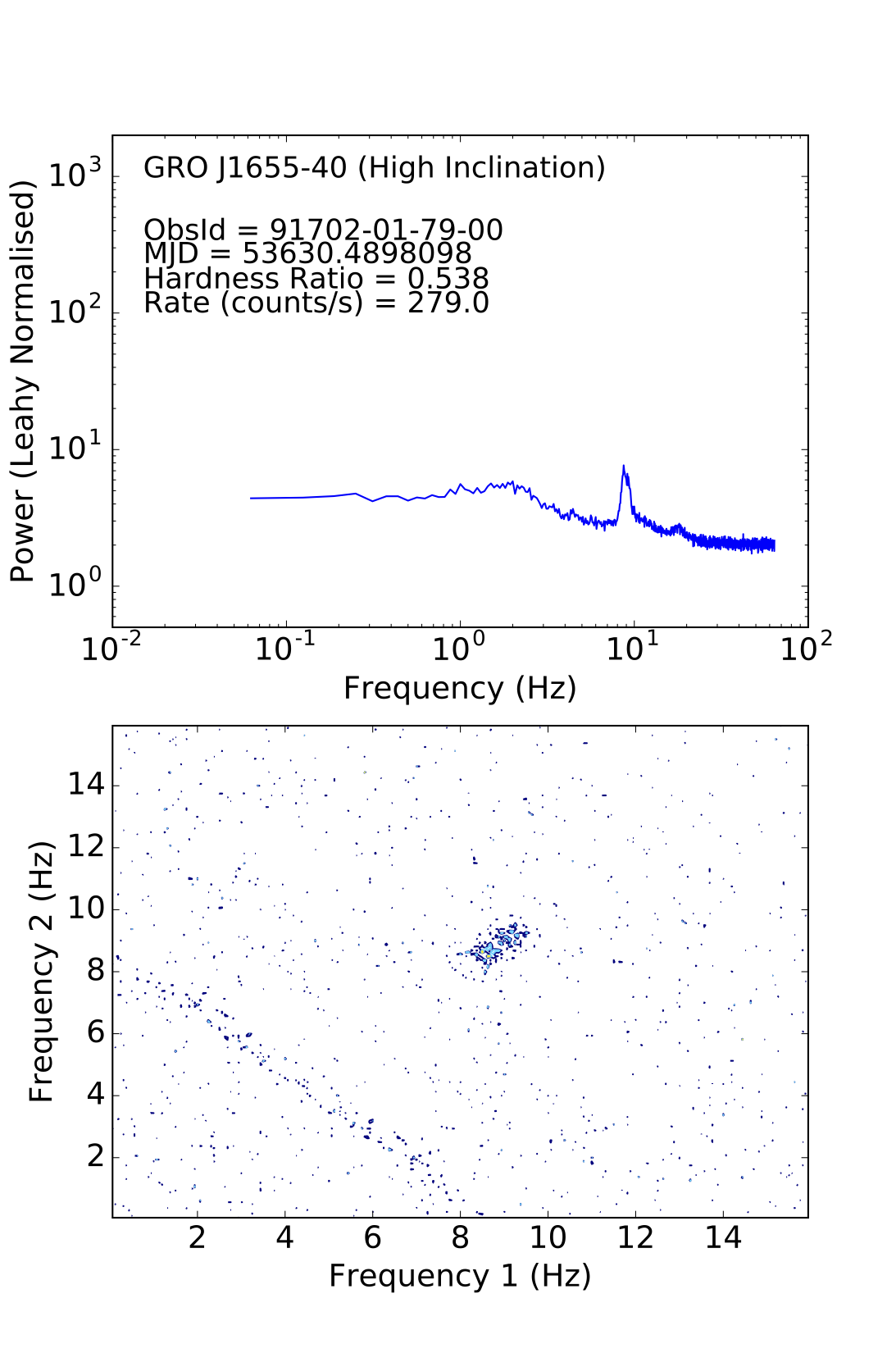}
    \caption{The power spectrum (Top panel) and bicoherence plot (Bottom panel) showing the 'hypotenuse' pattern during the hardening phase of an outburst from the high inclination source GRO~J1655-40 (observation 91702-01-79-00). The colour scheme of log$b^2$ is as follows: dark blue:-2.0, light blue:-1.75, yellow:-1.50, red:-1.25}
    \label{fig:hardening}
\end{figure}

\subsubsection{Intermediate inclination sources}

XTE~J1859+226 and MAXI~J1543-564 could not be unambiguously placed into either the low inclination or high inclination categories. All five observations of type C QPOs from MAXI~J1543-564 have low count rates (<100 counts/sec), and thus reliable bicoherence measurements could not be obtained. However, extensive coverage of the 1999 outburst from XTE~1859+226 was available for our analysis. We find that this source shows a transition from `web' to `cross' pattern, consistent with the behaviour of high inclination sources. This is also consistent with previous findings of \citep{Motta2015} where the QPO and noise rms properties of XTE~J1859+226 and of \citep{VandenEijnden2017} where the relative difference between the QPO centroid frequencies in different energy bands and the phase lag behaviour indicate that XTE~J1859+226 is a high inclination source.

\subsubsection{Quantifying the inclination dependence}

\begin{figure*}
	\includegraphics[width=\textwidth]{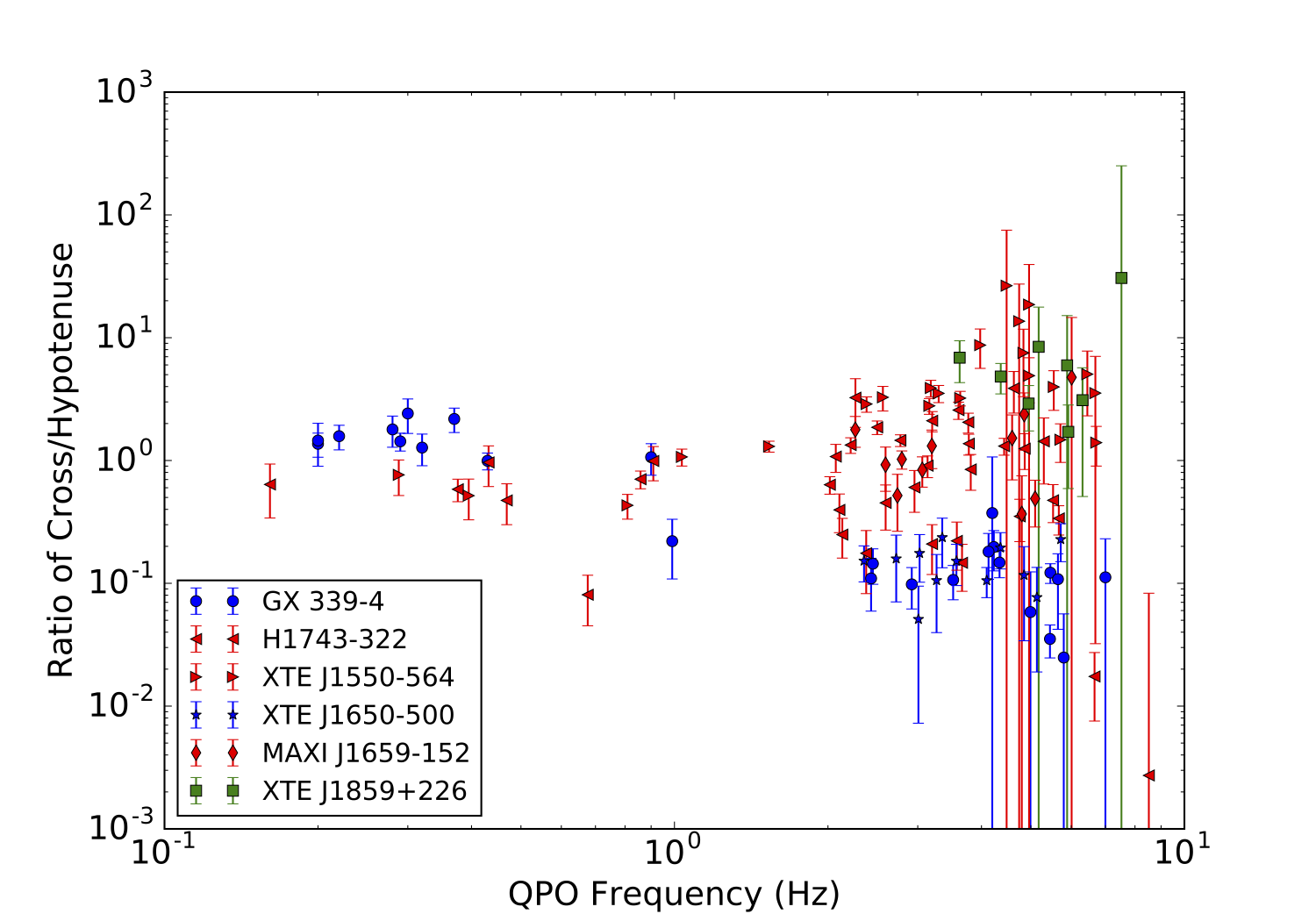}
    \caption{The ratio between the mean value of the bicoherence along the `hypotenuse' and `cross' as a function of QPO frequency. The green squares indicate the source XTE~J1859+226 which is of unknown inclination, but shows a behaviour consistent with high inclination sources. A colour version of this figure is available in the online version of this paper. }
    \label{fig:ratio}
\end{figure*}

The inclination dependence of this change is illustrated in Fig.~\ref{fig:ratio}, where we plot the ratio of mean values of the bicoherence along the hypotenuse and cross as a function of QPO frequency. In this plot we include the sources that have a reliable bicoherence estimates during the softening phase of the transition over a range of QPO frequencies. This includes GX~339-4 and XTE~J1650-500 for the low inclination sample (blue points) and H1743-322, XTE~J1550-564 and MAXI~J1659-152 for the high inclination sample (red points). 

For the estimation of the bicoherence along the `cross', regions where 1.3$f_{QPO}$ > $f$ > 1.8$f_{QPO}$ and where 1.3$f_{2QPO}$ > $f$ > 1.8$f_{2QPO}$ were excluded, as these regions would be dominated by the coupling between the QPO and the harmonic. Similarly, when estimating the bicoherence along the `hypotenuse' the region around $f_1$=$f_2$=0.5$f_{QPO}$ was excluded for observations where the QPO frequency is greater than 2Hz to avoid regions dominated by coupling between the QPO and the subharmonic.  It can be seen in Fig.~\ref{fig:ratio} that this ratio decreases in low inclination sources above 2Hz, indicating a strengthening `hypotenuse' and a weakening `cross'. The opposite behaviour is seen in the high inclination sources, indicating a strengthening `cross' and a weakening `hypotenuse'. XTE~J1859+226, which is of unknown inclination (green points) shows a high ratio of cross to hypotenuse, similar to that of the high inclination sources. 

\subsection{Statistical Significance}

While a distinct difference can be seen in the behaviour between high and low inclination sources, the sample size of the sources is small. In this section, we investigate the possibility that the behaviour of the bicoherence is specific to the source and the inclination dependence has arisen from random chance. Thus our null hypothesis is that the behaviour of the bicoherence does not depend on the inclination angle of the source.

First, we consider the sample of 3 high inclination (H1743-322, XTE~J1550-564 and MAXI~J1659-152) and 2 low inclination (GX~339-4 and XTE~J1650-500) sources. Assuming that each source has a 50\% chance of falling into either behaviour category, using a binomial distribution gives a p-value of p=0.0625 that the behaviour coincides with source inclination.

Based on the QPO and noise rms properties and the phase lag behaviour, if XTE~J1859+226 is included in our sample as a high inclination source the p-value now becomes p=0.03125. These values are strongly suggestive of an underlying inclination dependence, especially when considering the inclination dependence of other variability properties. However, it is also possible that any individual source is more likely to exhibit one bicoherence behaviour over the other. Thus the addition of more sources is required to make a more definite claim on the statistical significance of the inclination dependence.

\subsection{Type B QPOs}

\begin{figure}
	\includegraphics[width=\columnwidth]{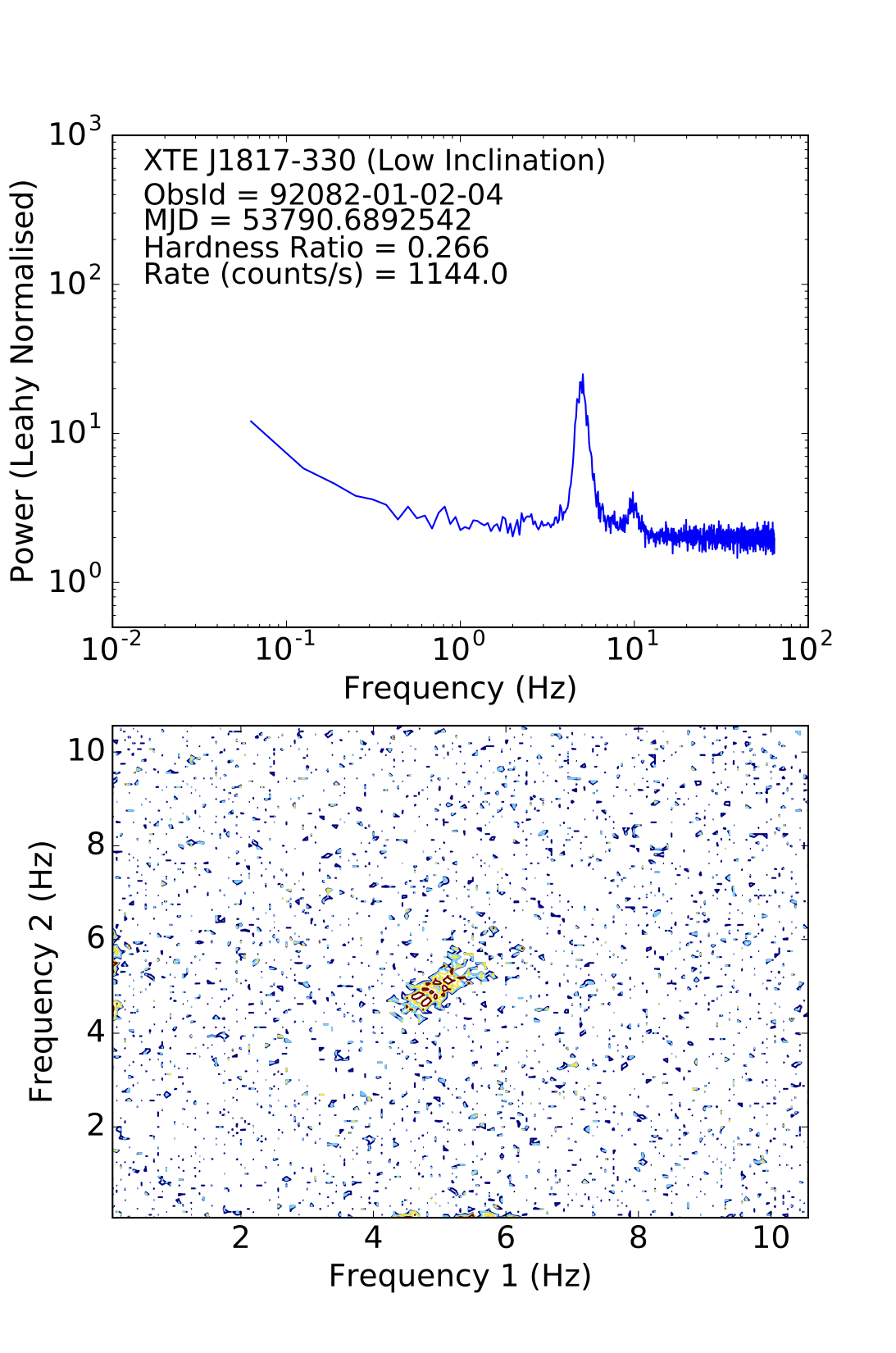}
    \caption{The power spectrum (Top panel) and bicoherence plot (Bottom panel) showing the typical pattern seen from a Type B QPO. The plot is from the low inclination source XTE~J1817-330 (observation 92082-01-02-04). The colour scheme of log$b^2$ is as follows: dark blue:-2.0, light blue:-1.75, yellow:-1.50, red:-1.25}
    \label{fig:typeb}
\end{figure}

Almost all of the observations of type B QPOs show a high bicoherence in the region where both $f_1$ and $f_2$ are equal to $f_{QPO}$, indicating the presence of a second harmonic (often visible in the power spectrum). The presence of a sub harmonic is also occasionally indicated by a feature in the bicoherence plot where $f_1$ and $f_2$ are equal to half of $f_{QPO}$. However, no coupling is seen between the QPO and the broadband noise frequencies. Whether this is due to the weakness of the broadband noise in the SIMS, or a true lack of coupling as a result of type-B QPOs arising from a physically different mechanism is presently unclear. An example of the bicoherence from a Type B QPO is shown in Fig.~\ref{fig:typeb}. Additionally, unlike type C QPOs, the type B QPOs do not show any measurable inclination dependence in their bicoherence patterns. 

\section{Discussion}
\label{sec:discussion}

In this section, we expand on the model of a moderate increase in the optical depth of the corona as outlined in \citet{Arur2019} to explain the inclination dependence in the gradual change of the bicoherence. While the results are model independent, we first consider an interpretation in the context of the Lense-Thirring precession model with the optical depth varying over the precession timescales, as this model is well developed and easy to apply simple prescriptions to.

\subsection{Low inclination}
\label{sec:li_disc}

As described in \cite{Arur2019}, an increase in the optical depth of the corona (from $\tau$ $\sim$ 1 to $\tau$ $\sim$ 4) could lead to the bicoherence pattern gradually changing from a `web' to a `hypotenuse' pattern. In this interpretation, the initial `web' pattern seen in the HIMS is a combination of two effects: 1. The low frequency variability originating from the outer regions of the accretion disk modulate the QPO as the soft photons from the thin disk are upscattered by the corona. This produces the diagonal region of high bicoherence, which is produced by amplitude modulation. 2. The precession of the inner accretion flow modulating the high frequency variability that originates from the inner region of the accretion disk. This produces the vertical and horizontal regions of high bicoherence.  

As the source moves from the HIMS to SIMS, the QPO frequency increases and the spectrum of the source is dominated by the disc black body component. As there is a larger number of seed photons available for inverse Compton scattering by the corona, the `hypotenuse' becomes stronger in this state. On the other hand, an increase in the optical depth of the corona results in the high frequency variations being smeared out as the time scales for the diffusion of the photons through the corona is larger than the time scale of the variation. This results in the horizontal and vertical stripes having lower bicoherence as the phases of the HF variability and the QPO are no longer coupled. Overall, this results in the bicoherence moving from a `web' to a `hypotenuse' pattern as seen in Fig.~\ref{fig:li_evolution}.

\subsection{High inclination}

In the case of high inclination sources, the initial detection of a `web' pattern is likely due to the combination of the same two effects described in Section~\ref{sec:li_disc}. 

\begin{figure*}
	\includegraphics[width=0.4\textwidth]{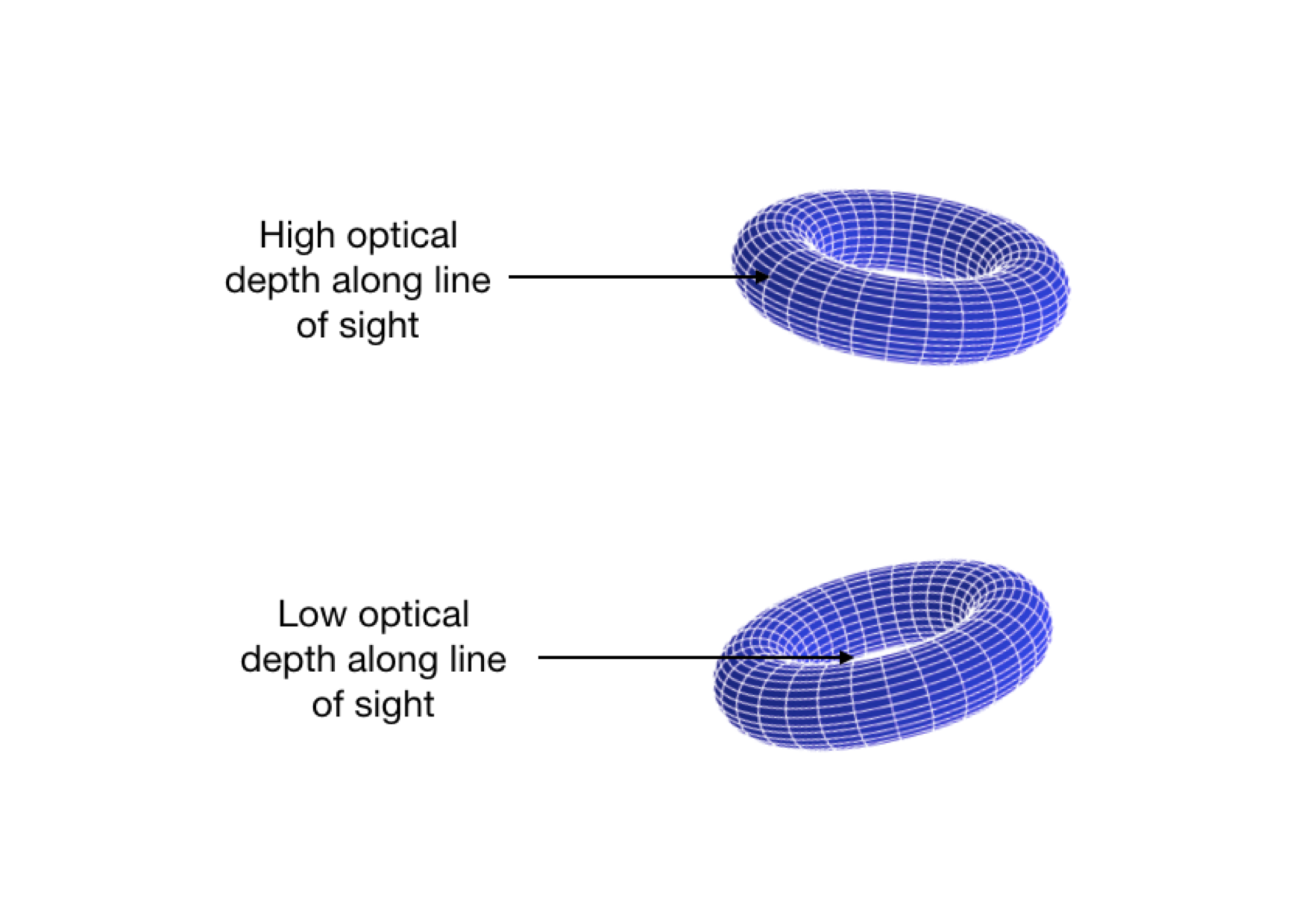}
	\includegraphics[width=0.4\textwidth]{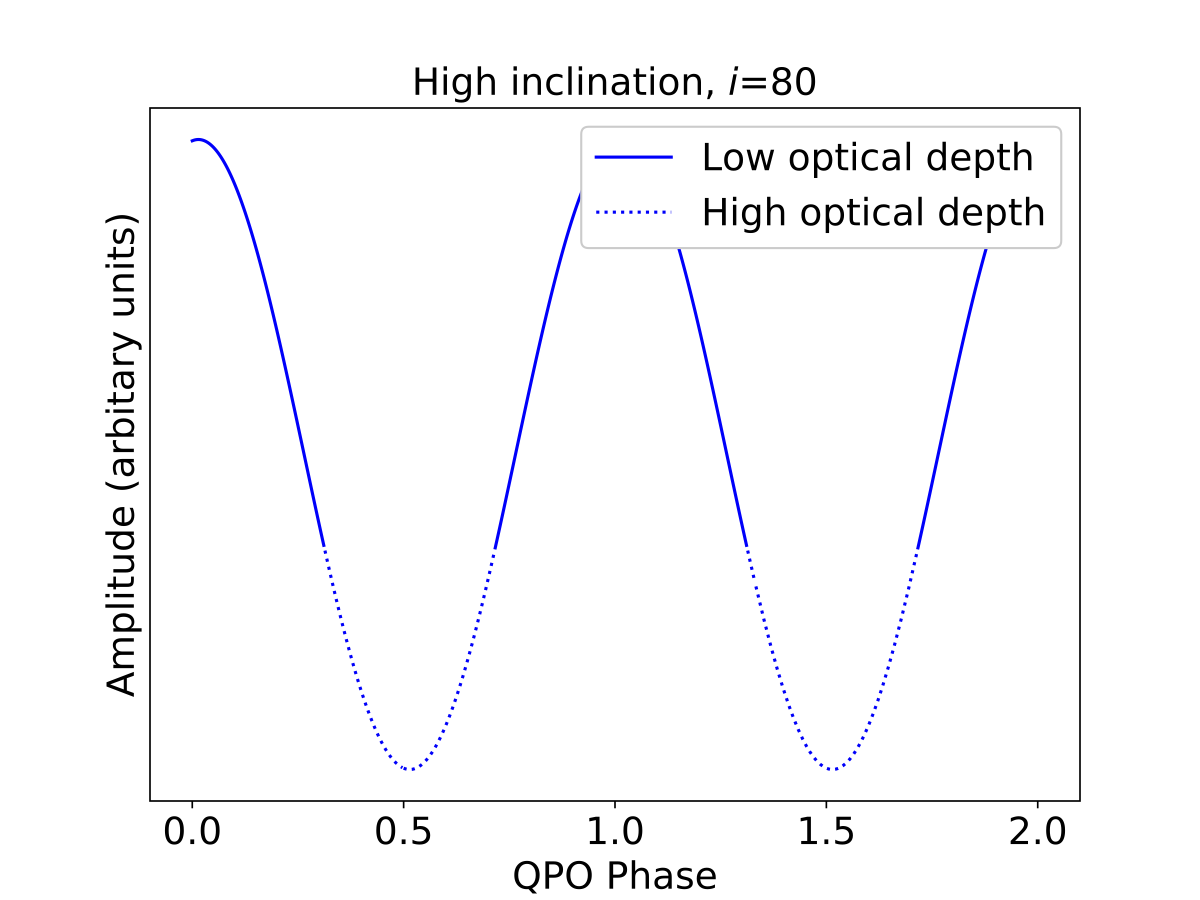}
    \caption{Left Panel: An illustration of the variation in the optical depth along the line of sight for different QPO phases for high inclination sources. Right Panel: A plot of the variation of the optical depth with QPO phase. The dotted line indicates the phases where the optical depth is high, and the solid line indicates the phases where it is low for a torus of $H/R$=0.3. A misalignment angle of 10$^{\circ}$, azimuthal angle of 5$^{\circ}$ and an inclination angle of 80$^{\circ}$ is assumed. In this case, the optical depth varies with the QPO phase. }
    \label{fig:torus_hi}
\end{figure*}

However, as the source transitions to the HIMS, the bicoherence pattern gradually changes to a `cross' pattern. Such a change is possible if the shape of the optically thick corona is radially extended with a low scale height (e.g as in a toroidal region). In this scenario, the optically thick ($\tau$ $\sim$ 4) corona is viewed at high inclination (i.e edge on). As this region precesses, the optical depth along the line of sight to the observer is either low or high depending on the phase of the precession (see Fig~\ref{fig:torus_hi}). When the torus is viewed edge-on, the high frequency variability is scattered out due to the higher $\tau$ along the line of sight. However, when the torus is viewed more face-on, the $\tau$ along the line of sight is lower, and the high frequency variability is not scattered out. This causes the high frequency variability to be strongly coupled to the QPO, leading to the prominent `cross' pattern that is observed. 

While the spectrum is dominated by the disc black body component, the diagonal `hypotenuse' component is not observed in the high inclination sources during the softening phase. This can be explained in the above scenario if the optical thick corona is radially extended. In this case, due to the high optical depth, the increase in the X-ray luminosity in response to an increase in the mass accretion rate is delayed due to multiple scattering events. If the timescale of the scatterings is longer than the timescale of the accretion rate fluctuations, this results in a diluted coupling between the two, leading to a loss of coherence and the lack of a `hypotenuse' feature. This scenario also explains the brief emergence of the `hypotenuse' pattern in the high inclination sources, as the optical depth of the Comptonising corona drops in the hardening phase of the outburst. 

\subsection{Variation of the optical depth with QPO phase}

To provide a more quantitative illustration of the scenario described in the previous section, we assume the geometry and coordinate system described in \citet{Ingram2015a}.

\begin{figure*}
	\includegraphics[width=0.4\textwidth]{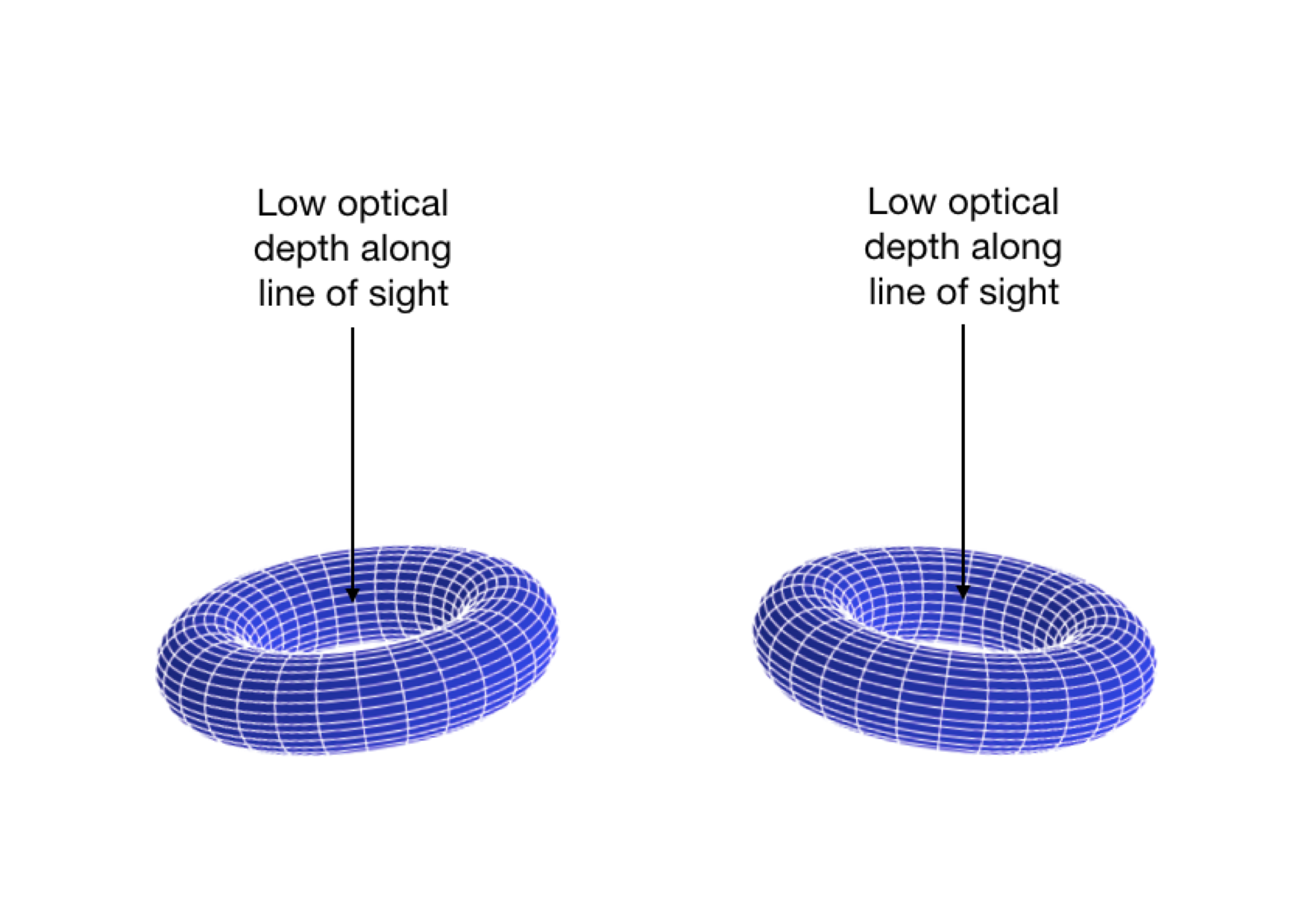}
	\includegraphics[width=0.4\textwidth]{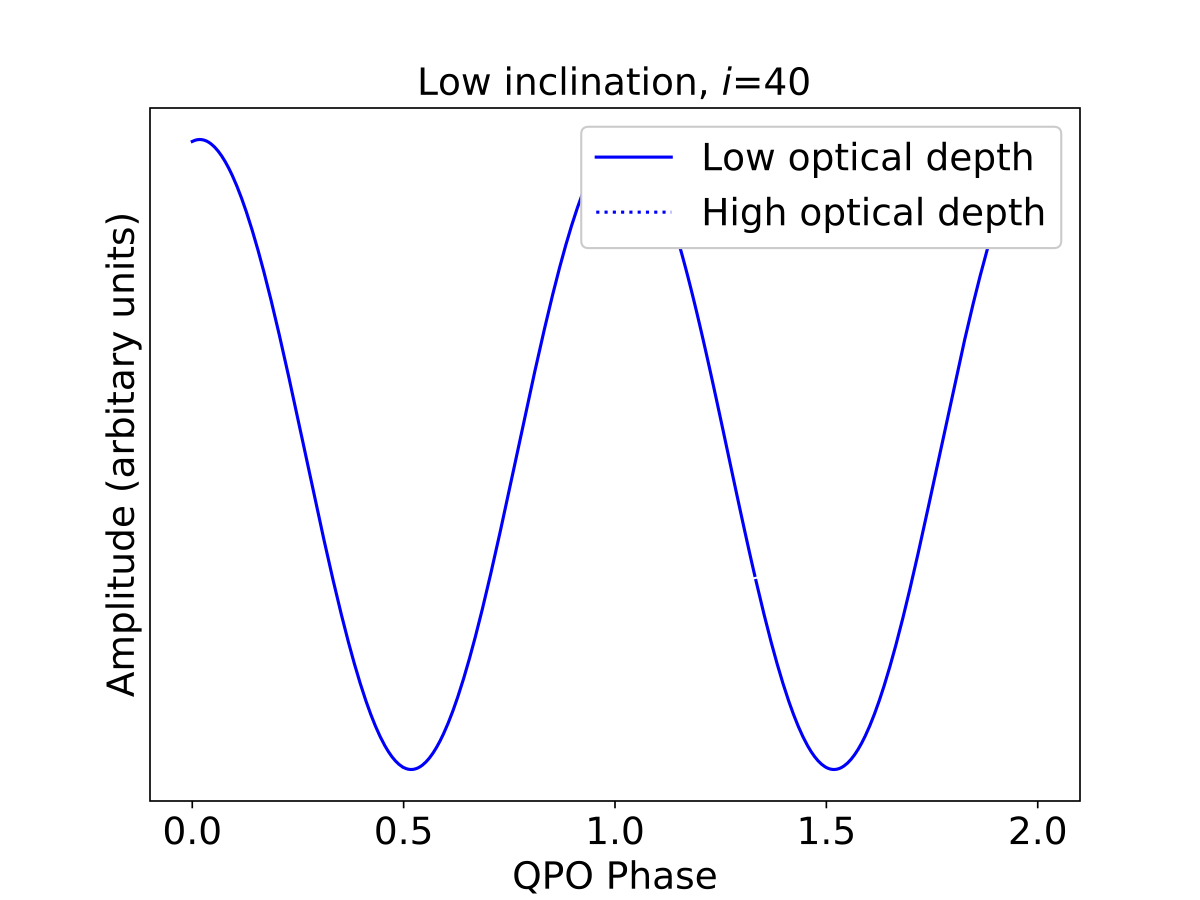}
    \caption{Left Panel: An illustration of the variation in the optical depth along the line of sight for different QPO phases for low inclination sources. Right Panel: A plot of the variation of the optical depth with QPO phase. The dotted line indicates the phases where the optical depth is high, and the solid line indicates the phases where it is low for a torus of $H/R$=0.3. A misalignment angle of 10$^{\circ}$, azimuthal angle of 5$^{\circ}$ and an inclination angle of 40$^{\circ}$ is assumed. In this case, the optical depth remains low throughout the QPO cycle.}
    \label{fig:torus_li}
\end{figure*}

Here, the BH spin is misaligned with the spin of the binary by an angle $\beta$. The `binary' coordinate system has $z$-axis $\bm{{\hat{z}_b}}$ that aligns with the orbital motions axis of the binary. In these binary coordinates the vector pointing from the BH to the observer is given by 

\begin{equation}
    \hat{\mathbf{o}} = (\text{sin } i \text{ cos } \Phi , \text{ sin }  i \text{ sin } \Phi , \text{ cos } i)
\end{equation}

where $i$ is the inclination angle and $\Phi$ is the azimuth of the observer. Assuming that the $z$-axis of the accretion flow $\bm{{\hat{z}_f}}$ precesses around the BH spin axis, maintaining a misalignment angle of $\beta$ as the precession angle $\omega$ increases. Thus in the `BH' coordinate system, the observers line of sight can be written as

\begin{equation}
    \hat{\mathbf{o}} = (\text{sin } \theta_0 , \text{ 0 } , \text{ cos }  \theta_0)
\end{equation}

where cos $\theta_0$, the angle between the observer and the BH spin axis, is given by:

\begin{equation}
    \text{cos }  \theta_0 = \text{ sin } i \text{ cos } \Phi \text{ sin } \beta + \text{ cos } i \text{ cos } \beta
\end{equation}

Thus the $z$-axis of the flow $\bm{{\hat{z}_b}}$ can be written in this coordinate system as:

\begin{equation}
    \bm{{\hat{z}_f}} = (\text{ sin }\beta \text{ cos }(\omega - \omega_0), \text{ sin }\beta \text{ sin }(\omega - \omega_0), \text{ cos} \beta )
\end{equation}

where $\omega_0$ is the precession angle at which the projection of $\bm{{\hat{z}_f}}$ onto the BH equatorial plane aligns with the BH $x$-axis (see Fig.1 of \citet{Ingram2015a}). $\omega_0$ is given by

\begin{equation}
    \text{tan }\omega_0 = \frac{\text{ sin } i \text{ sin } \Phi}{\text{ sin } i \text{ cos } \Phi \text{ cos } \beta - \text{ cos } i \text{ sin } \beta}
\end{equation}

The projection of the flow $z$-axis $\bm{{\hat{z}_f}}$ onto the observer line of sight as it precesses is plotted in Fig.\ref{fig:torus_hi} for high inclination and in Fig.\ref{fig:torus_li} for low inclination sources. \\

We also assume a simple model for an accretion flow such that if the angle between the observer and the line perpendicular to $\bm{{\hat{z}_f}}$ (the `viewing angle') is greater than a value $\psi$, the optical depth along the line of sight is low. If the viewing angle is lower than $\psi$, the optical depth along the line of sight is high. $\psi$ is given by the inverse tangent of the $H/R$. It can be seen in Fig.~\ref{fig:torus_hi} that for high inclination example, the optical depth along the line of sight varies with QPO phase. However, for the low inclination example, the optical depth along the line of sight remains low throughout the QPO cycle. \\

\subsection{Optical depth and geometry of the corona}

Oscillations of frequency $f$ are shown to be smeared by electron scattering when in a spherical cloud of radius R with an optical depth of $\tau$ if $2\pi f \tau R/c \gg 1$ \citep{Kylafis1987}. Thus, it is possible to estimate a lower limit for the optical depth along the different lines of sight, based on the frequencies involved in the scenario outlined above. \\

For the low inclination sources, the high frequency ($\sim$ 10Hz) variability of the `cross' pattern becomes obscured. Assuming a 10$M_{\odot}$ black hole with a comptonising region of 100 gravitational radii, an optical depth of $\tau \gg 3.2$ is required when the corona is viewed face on. \\

In the case of high inclination sources, the low frequency ($\sim$ 1Hz) variability of the `hypotenuse' pattern is obscured. This requires an optical depth of $\tau \gg 32$ when the corona is viewed edge on. This combination suggests that the corona is radially extended, with $H/R$ $\sim$ 0.3 (since the optical depth increases exponentially with radius). \\

To investigate the value of $\tau$ for the proposed scenario, here we estimate the optical depth of a radially extended torus assuming an alpha disk model \citep{Shakura1973}. The density of the torus with major radius R can be estimated to be: 

\begin{equation}
    \rho(R) = \frac{\dot{M}t_{visc}}{\pi^2 R^3}
\end{equation}

where the viscous timescale $t_{visc}\sim R^2/\alpha c_s H$, with $H$ being the scale height. The sound speed $c_s \sim \sqrt{kT/m_p}$ where $m_p$ is the proton mass. \\

The optical depth $\tau = \sigma_T S / m_p $, where the radially integrated surface density $S$ is given by: \\

\begin{equation}
    S(R) = \int_{R_{in}}^{R_{out}} \! \frac{\dot{M}R^2}{\pi^2 \alpha c_s H R^3} \, \mathrm{d}R  \approx \frac{\dot{M}}{\pi^2 \alpha (H/R) c_s } \left[ \frac{1}{R_{in}}- \frac{1}{R_{out}} \right]
\end{equation} \\

Here we assume a value of 0.1 for $\alpha$ and $H/R$ = 0.3 as estimated above. A value of kT $\sim$ 130 keV was assumed as this was the high energy cut off measured during the intermediate state of GX~339-4 \citep{Motta2009}. \\

For a mass accretion rate of $10^{19}$ g/s and a value of $R_{in} = 10^7$m (with $R_{in} << R_{out})$, this yields a $\tau =$ 38.6 that is consistent with the scattering time scales required. However, it must be noted that this value for the inner radius of the accretion flow is much larger than those assumed in the Lense-Thirring models. For smaller inner radii such as $R_{in}$ = $3 \times 10^5$m ($\sim$ 5$R_g$), $\tau =$ 1287. At $R_{in}$ =$10^5$m ($\sim$ 1.5$R_g$), the optical depth becomes $\tau =$ 3860.  \\

It should also be emphasized that the optical depth depends on the details of geometry of the torus, and thus can easily vary by a factor of $\sim$3 by changing the extent to which the torus is radially extended. \\

\subsection{Caveats}

We emphasise that the calculations in the interpretation above have many simplifying assumptions, and more complete numerical simulations are needed in order to make detailed comparisons between the model and observations. \\

While the calculations above are based on the the model of \citet{Kylafis1987}, the geometry assumed for the scattering timescales is based on a source of the seed photos that is embedded in a spherical Comptonising cloud. In the case of a radially extended toroidal structure illuminated by seed photons from a disk, the scattering time scales are likely to be much shorter. \\

The optical depth estimates are made under the assumption of a constant sound speed and a constant $H/R$ throughout the region. However, the scale height of the disk $H$ is set the balance between vertical pressure and gravity, and thus depends on the sound speed such that $H=c_s/\Omega_k$ where $\Omega_k$ is the Keplerian angular velocity. In this case

\begin{equation}
    \frac{H}{R} = \frac{c_s}{R \sqrt{\frac{GM}{R^3}}} = \frac{c_s}{c} \sqrt{\frac{R}{R_g}}
\end{equation}

For the sound speed at kT = 130keV, the value of H/R equals 0.3 at R $\approx 10^7$m, and becomes thinner at small radii. Additionally, it is likely that thicker accretion flows require a 2 temperature plasma, where the protons are at a much higher temperature than the inferred electron temperature.

Changes in the optical depth will not have one-to-one correspondences with changes in the fitted source spectra, unless the corona is single temperature, has a purely thermal electron distribution and maintains a specific radial density profile.  In the brightest hard states and in intermediate states, where these quasi-periodic oscillations are most often seen, it is likely that the corona has a range of temperatures and it is not clear how its geometry may be changing with time. Furthermore, many observations in these states which extend into the soft gamma-rays show non-thermal tails \citep{McConnell2002}. However, the observed spectra provide useful constraints on the optical depth. In the cases where extremely high optical depths are predicted, the spectra would tend towards that of a blackbody, in contrast with the observations. If a broad range of temperatures is present and/or the electron energy distribution includes a substantial non-thermal component, higher optical depths can be accommodated while still allowing the spectrum not to look like a blackbody, but as $\tau$ approaches hundreds, as is needed to have scattering-induced light travel time delays produce the relationships seen here if the coronal size is small, then the spectrum must tend toward looking like a blackbody corresponding to the outer region's temperature; the model proposed is likely viable only if the coronal size is at the upper end of the possible range and the range of temperatures is quite large.

In this paper, we have focused our discussion of the bispectral analysis on its meaning in the context of the Lense-Thirring precession model, but the observational results are model-independent. We emphasize that interpretation is presented as an outline to motivate future detailed modelling efforts to explain the non-linear behaviour of QPOs. It is possible that other models could also produce such non-linearity. For example, in the accretion ejection instability (AEI) model, the harmonics are produced are a result of general relativistic effects and show an inclination dependence \citep{Varniere2016a}. Also in this model, the QPO is linked to the band limited noise, making it possible to reproduce the hypotenuse pattern (P. Varni\`ere, private communication). However, due to computational expenses involved in the detailed numerical simulations of this model,  analytic approximations are additionally not straightforward, and current numerical simulations do not have data with the combination of sufficient length and time resolution to probe the nonlinearity quantitatively, so future numerical work will be required to compare the AEI model to the bispectral data.

\section{Conclusions}
\label{sec:conclusions}

We have presented the first systematic analysis of the inclination dependence of the non-linear properties of QPOs from BHXBs. We find that:

\begin{enumerate}

    \item Type C QPOs from high inclination sources show a gradual change from `web' to `cross' bicoherence patterns.

    \item Type C QPOs from low inclination sources show a gradual change from `web' to `hypotenuse' bicoherence patterns.

    \item The evolution of Type C QPOs from the intermediate inclination source XTE~J1859+226 is consistent with our sample of high inclination sources.
    
    \item Type-B QPOs do not show any inclination dependence in their bicoherence patterns. 

\end{enumerate}

We also propose a scenario of an increase in the optical depth of a radially extended corona to explain the change in the bicoherence in both the high and low inclination sources.

\section*{Acknowledgements}

The authors would like to thank Adam Ingram and Peggy Varni\`ere for helpful and interesting discussions. We would also like to thank the anonymous referee for a detailed and constructive report that greatly improved the content and clarity of the paper.

%%%%%%%%%%%%%%%%%%%%%%%%%%%%%%%%%%%%%%%%%%%%%%%%%%



\bibliographystyle{mnras}
\bibliography{Inclination} % if your bibtex file is called example.bib

%%%%%%%%%%%%%%%%%%%%%%%%%%%%%%%%%%%%%%%%%%%%%%%%%%

%%%%%%%%%%%%%%%%% APPENDICES %%%%%%%%%%%%%%%%%%%%%

%%%%%%%%%%%%%%%%%%%%%%%%%%%%%%%%%%%%%%%%%%%%%%%%%%

% Don't change these lines
\bsp	% typesetting comment
\label{lastpage}
\end{document}